\documentclass[pdftex,a4paper,twocolumn,epjc3]{svjour3}
\usepackage[english]{babel}
\usepackage[utf8]{inputenc}
\usepackage[T1]{fontenc}

\smartqed

\usepackage{mathptmx} 
\usepackage{amsmath}
\usepackage[pdftex]{graphicx}
\usepackage{xspace}
\usepackage{multirow}
\usepackage{enumerate}
\usepackage{flushend}
\usepackage[numbers,sort&compress]{natbib}
\usepackage[colorlinks,citecolor=blue,urlcolor=blue,linkcolor=blue]{hyperref}

\journalname{Eur. Phys. J. C}

\newcommand{\eV}{\ensuremath{\mbox{e\kern-0.1em V}}\xspace}
\newcommand{\GeV}{\ensuremath{\mbox{Ge\kern-0.1em V}}\xspace}
\newcommand{\MeV}{\ensuremath{\mbox{Me\kern-0.1em V}}\xspace}
\newcommand{\GeVc}{\ensuremath{\mbox{Ge\kern-0.1em V}\!/\!c}\xspace}
\newcommand{\GeVcc}{\ensuremath{\mbox{Ge\kern-0.1em V}\!/\!c^2}\xspace}
\newcommand{\AGeV}{\ensuremath{A\,\mbox{Ge\kern-0.1em V}}\xspace}
\newcommand{\AGeVc}{\ensuremath{A\,\mbox{Ge\kern-0.1em V}\!/\!c}\xspace}
\newcommand{\MeVc}{\ensuremath{\mbox{Me\kern-0.1em V}/c}\xspace}

\newcommand{\dedx}{\ensuremath{{\rm d}E\!/\!{\rm d}x}\xspace}

\DeclareMathOperator{\atanh}{atanh}

\newcommand{\Urqmd}{{\scshape U}r{\scshape qmd}\xspace}

\newcommand{\VenusLong}{{\scshape Venus4.12}\xspace}

\newcommand{\Epos}{{\scshape Epos}\xspace}
\newcommand{\EposLong}{{\scshape Epos1.99}\xspace}

\newcommand{\NASixtyOne}{NA61\slash SHINE\xspace}
\newcommand{\CernVM}{\textsc{Cern\-\kern-0.05emVM}\xspace}

\begin{document}

  \title{Multiplicity and transverse momentum fluctuations in
         inelastic proton-proton interactions at the CERN Super Proton Synchrotron
    }

  \institute{
    {National Nuclear Research Center, Baku, Azerbaijan}\label{inst0}
\and{Faculty of Physics, University of Sofia, Sofia, Bulgaria}\label{inst1}
\and{Ru{\dj}er Bo\v{s}kovi\'c Institute, Zagreb, Croatia}\label{inst2}
\and{LPNHE, University of Paris VI and VII, Paris, France}\label{inst3}
\and{Karlsruhe Institute of Technology, Karlsruhe, Germany}\label{inst4}
\and{Fachhochschule Frankfurt, Frankfurt, Germany}\label{inst5}
\and{University of Frankfurt, Frankfurt, Germany}\label{inst6}
\and{University of Athens, Athens, Greece}\label{inst7}
\and{Wigner Research Centre for Physics of the Hungarian Academy of Sciences, Budapest, Hungary}\label{inst8}
\and{Institute for Particle and Nuclear Studies, KEK, Tsukuba, Japan}\label{inst9}
\and{University of Bergen, Bergen, Norway}\label{inst10}
\and{Jan Kochanowski University in Kielce, Poland}\label{inst11}
\and{National Center for Nuclear Research, Warsaw, Poland}\label{inst12}
\and{Jagiellonian University, Cracow, Poland}\label{inst13}
\and{University of Silesia, Katowice, Poland}\label{inst14}
\and{Faculty of Physics, University of Warsaw, Warsaw, Poland}\label{inst15}
\and{University of Wroc{\l}aw,  Wroc{\l}aw, Poland}\label{inst16}
\and{Warsaw University of Technology, Warsaw, Poland}\label{inst17}
\and{Institute for Nuclear Research, Moscow, Russia}\label{inst18}
\and{Joint Institute for Nuclear Research, Dubna, Russia}\label{inst19}
\and{St. Petersburg State University, St. Petersburg, Russia}\label{inst21}
\and{University of Belgrade, Belgrade, Serbia}\label{inst22}
\and{ETH Z\"urich, Z\"urich, Switzerland}\label{inst23}
\and{University of Bern, Bern, Switzerland}\label{inst24}
\and{University of Geneva, Geneva, Switzerland}\label{inst25}
}

\author{
{A.~Aduszkiewicz}\thanksref{inst15}
\and{Y.~Ali}\thanksref{inst13,Ali}
\and{E.~Andronov}\thanksref{inst21}
\and{T.~Anti\'ci\'c}\thanksref{inst2}
\and{N.~Antoniou}\thanksref{inst7}
\and{B.~Baatar}\thanksref{inst19}
\and{F.~Bay}\thanksref{inst23}
\and{A.~Blondel}\thanksref{inst25}
\and{J.~Bl\"umer}\thanksref{inst4}
\and{M.~Bogomilov}\thanksref{inst1}
% ...
}

\thankstext{Ali}{Now at COMSATS Institute of Information Technology, Islamabad, Pakistan}

  \date{Received: date / Accepted: date}

  \titlerunning{$p_T$ and N fluctuations in p+p interactions at the CERN SPS}
  \authorrunning{The NA61/SHINE Collaboration}

  \maketitle


  \begin{abstract}
  Measurements of
  multiplicity and transverse momentum fluctuations
  of charged particles were performed in
  inelastic p+p interactions at 20, 31, 40, 80 and 158~\GeVc beam momentum.
  Results for the scaled variance of the multiplicity
  distribution and for three strongly intensive measures of multiplicity and transverse
  momentum fluctuations
  $\Delta[P_{T},N]$, $\Sigma[P_{T},N]$ and $\Phi_{p_T}$ are presented.
  For the first time the results on fluctuations are fully
  corrected for experimental biases.

  The results on multiplicity and transverse momentum fluctuations significantly deviate
  from expectations for the independent particle production.
  They also depend on charges of selected hadrons.
  The string-resonance Monte Carlo models \Epos and \Urqmd do not describe
  the data.

  The scaled variance of multiplicity fluctuations is significantly higher in
  inelastic p+p interactions than in central Pb+Pb collisions measured
  by NA49 at the same energy per nucleon.
  This is in qualitative disagreement with the predictions of the
  Wounded Nucleon Model.
  Within the statistical framework the enhanced multiplicity
  fluctuations in inelastic p+p interactions can be interpreted as due to
  event-by-event fluctuations of the fireball energy and/or volume.

\end{abstract}

  \section{Introduction and motivation}
\label{sec:introduction}

This paper presents experimental results on event-by-event
fluctuations of multiplicities and transverse momenta of
charged particles produced in inelastic p+p interactions
at 20, 31, 40, 80 and 158\GeVc.
The measurements were performed by the multi-purpose
NA61/SHINE~\cite{shine,Abgrall:2014xwa} experiment at
the CERN Super Proton Synchrotron (SPS).
They are part of the strong interaction programme devoted to the study of the
properties of the onset of deconfinement and search for the
critical point of strongly interacting matter. Within this program
a two dimensional scan in collision energy and size of colliding nuclei
is in progress. Data on p+p, Be+Be and Ar+Sc collisions
were already recorded and data on p+Pb and Xe+La collisions will be registered within
the coming years. The expected signal of a critical point is a non-monotonic
dependence of various fluctuation measures in such a scan,
for recent review see Ref.~\cite{Gazdzicki:2015ska}.

The NA49 experiment~\cite{Afanasev:1999iu}
published results for
central Pb+Pb collisions in the collision energy range 20$A$ to 158\AGeV,
as well as
for p+p, C+C and Si+Si reactions at 158\AGeV.
Multiplicity fluctuations were measured in terms of the scaled variance
of the multiplicity distribution~\cite{Alt:2006jr,Alt:2007jq}
and fluctuations of the
transverse momentum of the particles were studied employing
measures $\Phi_{p_T}$~\cite{Anticic:2003fd,Anticic:2008aa}, recently
$\Delta[P_{T},N]$ and $\Sigma[P_{T},N]$ \cite{Anticic:2015fla}.

Also, at SPS energies results on event-by-event fluctuations
in Pb+Au collisions on mean transverse momentum were published
by the CERES experiment~\cite{Adamova:2003pz} and in Pb+Pb collisions by the WA98
collaboration on charged particle multiplicity, transverse energy~\cite{Collaboration:2011rsa}
as well as the ratio of the charged to neutral pion multiplicity~\cite{Aggarwal:2001aa}.

An interpretation of the experimental results on nucleus-nucleus
collisions relies to a large extent on a comparison with the
corresponding data on p+p and p+A interactions.
However, in particular the available data on fluctuations are sparse.
Suitable fluctuation measurements for p+p interactions only exist
at 158~GeV/c beam momentum~\cite{Alt:2006jr,Anticic:2003fd}. Moreover, fluctuation measurements
cannot be corrected in a model independent manner
for partial phase space acceptance. Thus all measurements of
the scan should be performed in the same phase space region.

In nucleus-nucleus reactions the impact parameter of the collisions
cannot be tightly controlled. This problem results in additional
unwanted contributions to fluctuations the effect of which needs
to be suppressed by employing so-called strongly intensive measures.
In addition to $\Phi_{p_T}$ two recently proposed strongly intensive
quantities $\Delta[P_{T},N]$ and $\Sigma[P_{T},N]$ are studied in
this publication.

The paper is organized as follows. In Sec.~\ref{sec:measures} strongly
intensive fluctuation measures are introduced. The experimental setup
is presented in Sec.~\ref{sec:facility}. Data processing and simulation
and the analysis procedure are described in Sec.~\ref{sec:reconstruction}.
and Sec.~\ref{sec:analysis}, respectively. Results are
presented and discussed in Sec.~\ref{sec:results} and compared to
model calculations in Sec.~\ref{sec:models}. A summary and
outlook in Sec.~\ref{sec:summary} closes the paper.

Across this paper
the pion rapidity is calculated in the collision centre of
mass system: $y_{\pi} = \atanh(\beta_L)$,
where $\beta_L = p_L/E$ is the longitudinal ($z$) component of the
velocity,
$p_L$ and $E$ are pion longitudinal momentum and energy given
in the collision centre of mass system.
The transverse component of the momentum is denoted as $p_T$ and
the transverse mass $m_T$ is defined as $m_T = \sqrt{m_{\pi}^2 + (cp_T)^2}$,
where $m_{\pi}$ is the charged pion mass in \GeV.
The azimuthal angle $\phi$ is the angle between transverse momentum vector and
the horizontal ($x$) axis.
The nucleon mass and collision energy per nucleon pair in the centre of mass
system are denoted as $m_\mathrm{N}$ and $\sqrt{s_\mathrm{NN}}$, respectively.
  \section{Fluctuation measures}\label{sec:measures}

\subsection{Intensive fluctuation measures}

Event quantities are called intensive if they do {\it not} depend on the
volume of the system within the grand canonical ensemble of statistical mechanics.
Examples are the mean transverse momentum of particles or ratios of particle
numbers in the events.
In contrast, so-called extensive quantities (for example the mean multiplicity
or the variance
of the multiplicity distribution) are proportional to the system
volume.
Within the Wounded Nucleon Model~\cite{Bialas:1976ed} intensive quantities are
those which are independent
of the number of wounded nucleons, and extensive ones those which are
proportional to the
number of wounded nucleons. The ratio of two extensive quantities is an
intensive
quantity~\cite{Gorenstein:2011vq}. Therefore, the scaled variance of a
quantity $A$
\begin{equation}\label{eq:omega}
  \omega[A] = \frac{Var(A)}{\langle A \rangle} = \frac{\langle A^2 \rangle -
  \langle A \rangle ^2}{\langle A \rangle}
\end{equation}
is an intensive measure. In fact, due to its intensity property, the scaled
variance ($\omega[N]$) of the distribution of multiplicity $N$ in the events
is widely used to quantify multiplicity fluctuations in high-energy heavy-ion experiments.

The scaled variance assumes the value $\omega[N] = 0$ for $N = const.$
and $\omega[N] = 1$ for
a Poisson multiplicity distribution.

\subsection{Strongly intensive fluctuation measures}

Unfortunately, the volume of the matter produced in heavy ion collisions
cannot be fixed
and changes significantly from one event to another.
Therefore, it is very important to be able to measure the properties of the
created matter independently of its
volume fluctuations.
The quantities which allow this are called {\it strongly intensive} measures.
They depend neither on the volume nor on the fluctuations of the volume.
Ratios of mean multiplicities are both intensive and strongly intensive
measures.
The situation is, however, much more difficult for the analysis of
fluctuations.
For example the scaled variance is an intensive but not strongly intensive
measure.

It was shown in Ref.~\cite{Gorenstein:2011vq}, that for certain combinations
of scaled variances,
terms dependent on the volume fluctuations cancel out. There are at least two
families of strongly intensive measures of two fluctuating extensive quantities
$A$ and $B$:
\begin{eqnarray}
  \label{eq:delta}
  \Delta[A,B] &=& \frac{1}{C_{\Delta}} \biggl[ \langle B \rangle \omega[A] -
        \langle A \rangle \omega[B] \biggr] \\
  \label{eq:sigma}
  \Sigma[A,B] &=& \frac{1}{C_{\Sigma}} \biggl[ \langle B \rangle \omega[A] +
        \langle A \rangle \omega[B] - 2 \bigl( \langle AB \rangle -
        \langle A \rangle \langle B \rangle \bigr) \biggr].
\end{eqnarray}
For the study of transverse momentum fluctuations one uses:
\begin{displaymath}
  A = P_{T} = \sum\limits_{i=1}^{N} p_{T_{i}}, \qquad B = N ,
\end{displaymath}
where $p_{T_{i}}$ is the modulus of the transverse momentum of particle $i$.

There is an important difference between $\Delta[P_{T},N]$ and
$\Sigma[P_{T},N]$.
Only the first two moments: $\langle P_{T} \rangle$, $\langle N \rangle$,
and $\langle P_{T}^2 \rangle$, $\langle N^2 \rangle$ are required to
calculate $\Delta[P_{T},N]$, whereas $\Sigma[P_{T},N]$ includes the correlation
term
$\langle P_{T}N \rangle - \langle P_{T} \rangle \langle N \rangle$.
Thus $\Delta[P_{T},N]$ and $\Sigma[P_{T},N]$ can be
sensitive to various physics effects in different ways.
In Ref.~\cite{Gorenstein:2011vq} strongly intensive quantities including
the correlation term are named the $\Sigma$ family, and those based only
on mean values and variances the $\Delta$ family.

Historically, the first proposed strongly intensive fluctuations measure
was $\Phi$~\cite{Gazdzicki:1992ri}. When applied to transverse momentum
fluctuations the measure is called $\Phi_{p_T}$. This has already been
used extensively by the NA49 experiment~\cite{Anticic:2003fd,Anticic:2008aa}.
The measure is a member of the $\Sigma$ family:
\begin{equation}\label{eq:phi}
  \Phi_{p_T} = \sqrt{\overline{p_T} \omega[p_T]}
  \left[\sqrt{\Sigma[P_{T},N]}-1\right].
\end{equation}
where $\overline{p_T}$ and $\omega[p_T]$ denote the average and scaled
variance of the inclusive $p_T$ distribution.

With the normalization proposed in Ref.~\cite{Gazdzicki:2013ana},
\begin{equation}
  \label{eq:normalization}
  C_{\Delta} = C_{\Sigma} = \langle N \rangle \omega[p_{T}],
\end{equation}
these measures
are dimensionless and have a common scale required for a quantitative
comparison of fluctuations of different,
in general dimensional, extensive quantities. More precisely, the values of
$\Delta$ and $\Sigma$ are equal to
zero in the absence of event-by-event fluctuations ($N = const.$, $P_{T} = const.$)
and equal to one for fluctuations given by
the independent particle production model (IPM)
\cite{Gazdzicki:2013ana,Gorenstein:2013nea}.
The values of $\Delta[P_{T},N]$ and $\Sigma[P_{T},N]$ have already been
determined in several models. The results of
the IPM, the Model of Independent Sources (MIS), source-by-source temperature
fluctuations
(example of MIS), event-by-event (global) temperature fluctuations, correlation
between average $p_T$ per event
and its multiplicity were published in Ref.~\cite{Gorenstein:2013nea}. The effects
of acceptance losses, efficiency losses, quantum (Bose-Einstein and
Fermi-Dirac) statistics and centrality dependence (UrQMD) were investigated in
Ref.~\cite{Gorenstein:2013iua}. Finally, the system size and
energy dependence in the UrQMD model was studied in
Ref.~\cite{Gazdzicki:2013ana}.
One of the conclusions (supported by the UrQMD calculations) is that the
$\Delta[P_{T},N]$, $\Sigma[P_{T},N]$, and $\Phi_{p_T}$ quantities measure
deviations
from the superposition model in different ways.
Therefore, in the analysis of experimental data a simultaneous measurement
of all three quantities is highly desirable.

\begin{table*}[!htb]
  \caption{
    Properties of $\Phi_{p_T}$, $\Delta[P_{T},N]$, and $\Sigma[P_{T},N]$
    in the absence of fluctuations, and
    in the Independent Particle Model (IPM) and the Model of Independent
    Sources (MIS) ($N_S$ denotes the number of sources).
  }
  \begin{center}
    \begin{tabular}{c|c|c|c|c}
      \noalign{\hrule height 0.03cm}\hline
      & unit & No fluctuations & IPM & MIS \cr
      \hline
      $\Phi_{p_T}$ & \MeVc & $\Phi_{p_T}=- \sqrt { \overline{p_T} \omega[p_T]}$
      & $\Phi_{p_T}=0$ & does not depend on $N_s$ and its fluctuations \cr
      $\Delta[P_{T},N]$ & dimensionless & $\Delta[P_{T},N]=0$ &
      $\Delta[P_{T},N]=1$ & does not depend on $N_s$ and its fluctuations \cr
      $\Sigma[P_{T},N]$ & dimensionless & $\Sigma[P_{T},N]=0$ &
      $\Sigma[P_{T},N]=1$ & does not depend on $N_s$ and its fluctuations \cr
      \hline
      $\omega[N]$ & dimensionless & $\omega[N]=0$ & $\omega[N]=1$ & does
      not depend on $N_s$ \cr
      \hline\noalign{\hrule height 0.03cm}
    \end{tabular}

  \end{center}

  \label{tbl:measures}
\end{table*}

A comparison
of the properties of these three measures within the IMP and MIS models
is shown in Table~\ref{tbl:measures}. If one finds, e.g.
$\Phi_{p_T}=10$ \MeVc one does not know whether this is a large or a small
effect, especially when the magnitudes
of $\Phi_{p_T}$ from several ''trivial'' effects (Bose-Einstein statistics,
resonance decays, etc.) are not
estimated. The situation is, however, different for $\Sigma[P_{T},N]$. If
one measures,
for example, $\Sigma[P_{T},N]=1.1$ this means that (for this specific
combination of moments)
one measures 10\% deviation from the IPM (fluctuations are 10\% larger than
in the IPM).
Therefore, the new measures $\Delta[P_{T},N]$ and $\Sigma[P_{T},N]$ have
the advantages of $\omega[N]$ but they
also preserve the advantage of $\Phi_{p_T}$, i.e. they are {\it strongly}
intensive measures of fluctuations.
  \section{Experimental facility}\label{sec:facility}

\subsection{The \NASixtyOne detector}\label{sec:detector}

\begin{figure*}[!htb]
  \centering
  \includegraphics[width=0.8\textwidth]{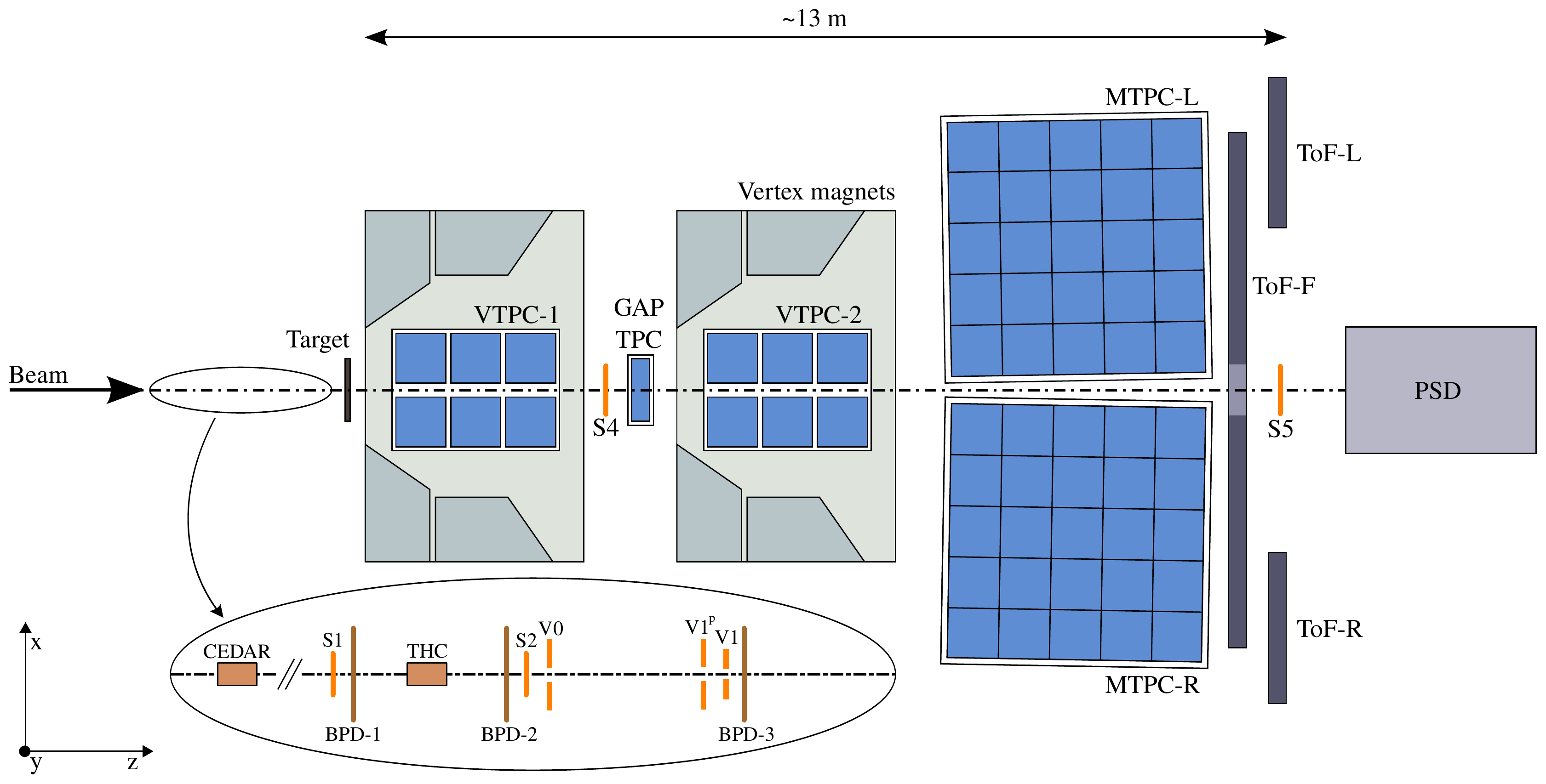}
  \caption[]{
  (Colour online)
  The schematic layout of the \NASixtyOne spectrometer
  (horizontal cut, not to scale).
  The beam and trigger detector configuration used for data
  taking in 2009 is shown in the inset.
  The chosen coordinate system is drawn on the lower left:
  its origin lies in the middle of the VTPC-2, on the beam axis.
  The nominal beam direction is along the $z$ axis.
  The magnetic field bends charged particle trajectories
  in the $x$--$z$ (horizontal) plane. Positively charged particles are bent
  towards the top of the plot.
  The drift direction in the TPCs is along the $y$ (vertical) axis.
  }
  \label{fig:detector}
\end{figure*}

The \NASixtyOne experimental facility~\cite{Abgrall:2014xwa}
consists of a large acceptance hadron
spectrometer located in the
CERN North Area Hall~887~(EHN1) and the H2 beam-line to which beams accelerated
in the CERN accelerator
complex are delivered from the Super Proton Synchrotron (SPS).
The schematic layout of the \NASixtyOne detector is shown in
Fig.~\ref{fig:detector}.

A set of scintillation and Cherenkov counters as well as beam position
detectors (BPDs) upstream
of the spectrometer provide timing reference, identification and position
measurements of incoming
beam particles.
Trajectories of individual beam particles were measured in a telescope of
beam position detectors placed
along the beam line (BPD-1/2/3 in Fig.~\ref{fig:detector}). These counters
are small ($4.8\times4.8$~cm$^2$)
proportional chambers with cathode strip readout, providing a resolution of
about 100~$\mu$m in two orthogonal
directions.  Due to properties
of the H2 beam line both
the beam width and divergence at the \NASixtyOne target increase with decreasing
beam momentum.
The trigger scintillator counter S4 placed downstream of
the target is used to
select events with collisions in the target area.
The liquid hydrogen target as well as the proton beams and triggers are
described in Secs.~\ref{sec:target} and~\ref{sec:beam}, respectively.

The main tracking devices of the spectrometer are four large volume Time
Projection Chambers (TPCs).
Two of them, the vertex TPCs (VTPC-1 and VTPC-2 in Fig.~\ref{fig:detector}),
are located in the
magnetic fields of two super-conducting dipole magnets with a maximum combined
bending power of 9~Tm which
corresponds to about 1.5~T and 1.1~T fields in the upstream and downstream
magnets, respectively.
In order to optimize the acceptance of the detector at each collision momentum,
the field in both
magnets was adjusted proportionally to the beam momentum.

Two large TPCs (MTPC-L and \mbox{MTPC-R}) are positioned downstream of the
magnets symmetrically
to the beam line. The fifth small TPC (\mbox{GAP-TPC}) is placed between
\mbox{VTPC-1} and \mbox{VTPC-2}
directly on the beam line. It closes the gap along the beam axis between the
sensitive volumes of the other TPCs.

The TPCs are filled with Ar:CO$_2$ gas mixtures in proportions 90:10 for
the VTPCs and the GAP-TPC,
and 95:5 for the MTPCs.

The particle identification capability of the TPCs based on measurements of
the specific energy loss, \dedx,
is augmented by time-of-flight measurements using Time-of-Flight (ToF)
detectors.
The high resolution forward calorimeter, the Projectile Spectator Detector
(PSD), measures energy
flow around the beam direction, which in nucleus-nucleus collisions is
primarily given by the
projectile spectators.

The results presented in this paper were obtained using information from
the TPCs
the Beam Position Detectors
as well as from the beam and trigger counters.

\subsection{Target}\label{sec:target}

\NASixtyOne uses various solid nuclear targets and a liquid hydrogen target
(see Sec.~\ref{sec:beam} for details).
For data taking on p+p interactions a liquid hydrogen target of 20.29~cm
length
(2.8\%~interaction length) and 3~cm diameter was
placed 88.4~cm upstream of VTPC-1.
The Liquid Hydrogen Target facility (LHT)
filled the target cell with para-hydrogen obtained
in a closed-loop liquefaction system
which was operated at 75~mbar overpressure with respect to the atmosphere.
At the atmospheric pressure of 965~mbar the liquid hydrogen density
is $\rho_\mathrm{LH} = 0.07$~g/cm$^3$.

Data taking with inserted ($I$) and removed ($R$) liquid hydrogen ($LH$)
in the LHT  was alternated in order to calculate a data-based correction for
interactions with the material surrounding the liquid hydrogen.

\subsection{Beams and triggers}\label{sec:beam}

Secondary beams of positively charged hadrons at 20, 31, 40, 80 and
158~\GeVc were produced from 400~\GeVc
protons extracted from the SPS onto a beryllium target in a slow extraction mode with a flat-top of
10 seconds.
The secondary beam momentum and intensity was adjusted by proper setting of
the H2 beam-line magnet currents and
collimators. The beam was transported along the H2 beam-line towards the
experiment.
The precision of the setting of the beam magnet currents was approximately
0.5\%. This was verified by a direct
measurement of the beam momentum at 31~\GeVc by bending the incoming beam
particles into the TPCs
with the maximum magnetic field~\cite{Abgrall:2011ae}. Selected beam
properties are given in Table~\ref{tab:beams}.

The set-up of beam detectors is illustrated in the inset on
Fig.~\ref{fig:detector}.
Protons from the secondary hadron beam were identified by two Cherenkov
counters, a CEDAR
(either \mbox{CEDAR-W} or \mbox{CEDAR-N}) and a threshold counter (THC).
The CEDAR counter,
using a coincidence of six out
of the eight photo-multipliers placed radially along the Cherenkov ring,
provided positive identification of
protons, while the THC, operated at pressure lower than the proton threshold,
was used in anti-coincidence in
the trigger logic. Due to their limited range of operation two different
CEDAR counters were employed,
namely for beams at 20, 31, and 40~\GeVc the \mbox{CEDAR-W} counter and
for beams at 80 and 158~\GeVc the \mbox{CEDAR-N} counter. The threshold counter
was used for 20, 31, and 40~\GeVc beams.
A selection based on signals from the Cherenkov counters allowed to identify
beam protons with a purity of
about 99\%. A consistent value for the purity was found by bending the 30.1~\GeVc beam
into
the TPCs with the full magnetic field and using the \dedx identification
method. The fraction of protons in the beams is given in
Table~\ref{tab:beams}.

\begin{table}[!htb]
  \caption{Basic properties of the beam used in the study of
           p+p interactions at 20, 31, 40, 80 and 158~\GeVc.
           The first column gives the beam momentum. The second and third
           columns list typical numbers of beam particles
           at \NASixtyOne per spill (about 10~seconds) and
           the fraction of protons in the beam,
           respectively.
}
\label{tab:beams}
\centering
  \begin{tabular}{c | c | c }
  \noalign{\hrule height 0.03cm}\hline
    $p_\mathrm{beam}$ [\GeVc] & particles per spill & proton fraction \\
    \hline
    20 & 1000k & 12\% \\
    31 & 1000k & 14\% \\
    40 & 1200k & 14\% \\
    80 & 460k & 28\% \\
    158 & 250k & 58\% \\
    \hline\noalign{\hrule height 0.03cm}
  \end{tabular}
\end{table}

Two scintillation counters, S1 and S2, provided beam definition, together
with the three veto
counters V0, V1 and V1$^\mathrm{p}$ with a 1~cm diameter hole, which were
defining the beam before the target.
The S1 counter also provided the timing (start time for the gating of all counters).
Beam protons were then selected by the coincidence:
\begin{equation}
T_{beam} =
\textrm{S1}
\wedge\textrm{S2}
\wedge\overline{\textrm{V0}}
\wedge\overline{\textrm{V1}}
\wedge\overline{\textrm{V1}^\mathrm{p}}
\wedge\textrm{CEDAR}
\wedge\overline{\textrm{THC}}~.
\end{equation}
The interaction trigger ($T_{int}$) was provided by the anti-coincidence of
the incoming proton beam and a scintillation counter S4
($T_{int} = T_{beam} \wedge\overline{\textrm{S4}}$).
The S4 counter with a 2\,cm diameter, was placed
between the VTPC-1 and VTPC-2 detectors along the  beam trajectory
at about 3.7~m from the target,
see Fig.~\ref{fig:detector}.
A large fraction of  beam protons that interact in the target does not reach S4.
The interaction and beam triggers were run simultaneously.
The beam trigger events were recorded with a frequency
by a factor of about 10
lower than the frequency of interaction trigger events.
  \section{Data processing and simulation}\label{sec:reconstruction}

Detector parameters were optimized by a data-based calibration procedure
which also took into account their
time dependence, for details see Refs.~\cite{Abgrall:2013pp_pim,Abgrall:1113279}.

The main steps of the data reconstruction procedure were:
\begin{enumerate}[(i)]
  \item cluster finding in the TPC raw data, calculation of the cluster
  centre-of-gravity and total charge,
  \item reconstruction of local track segments in each TPC separately,
  \item matching of track segments into global tracks,
  \item track fitting through the magnetic field and determination of track
  parameters at the first measured
        TPC cluster,
  \item determination of the interaction vertex using the beam trajectory
  ($x$ and $y$ coordinates) fitted
        in the BPDs and the trajectories of tracks reconstructed in the TPCs
        ($z$ coordinate),
  \item refitting the particle trajectory using the interaction vertex as
  an additional point and determining
        the particle momentum at the interaction vertex,
  \item matching of ToF hits with the TPC tracks.
\end{enumerate}

The accuracy of the transverse position of the main vertex is given by the resolution
of the BPDs ($\approx$100~$\mu$m). The resolution of the longitudinal position determination
is given by the TPC track reconstruction procedure and depends on the track multiplicity and magnetic field. For inelastic p+p interactions 158~\GeVc it is about 2~cm.

A simulation of the \NASixtyOne detector response was used to correct the
reconstructed data. Several MC
models were compared with the \NASixtyOne results on p+p, p+C and $\pi$+C
interactions: FLUKA2008,
URQMD1.3.1, VENUS4.12, EPOS1.99, GHEISHA2002, QGSJetII-3 and
Sibyll2.1~\cite{Abgrall:2011ae,ISVHECRI12_MU}.
Based on these comparisons and taking into account continuous support and
documentation from the developers
the EPOS model was selected for the MC simulation. The simulation consisted
of the following steps:
\begin{enumerate}[(i)]
  \item generation of inelastic p+p interactions using the EPOS model,
  \item propagation of outgoing particles through the detector material
  using the GEANT 3.21
        package which takes into account the magnetic field as well as
        relevant physics processes,
        such as particle interactions and decays,
  \item simulation of the detector response using dedicated \NASixtyOne
  packages which simulates charge clusters in the TPCs and introduces distortions
        corresponding to all corrections applied to the real data,
  \item simulation of the interaction trigger selection by checking whether
  a charged particle hits the S4
        counter, see Sec.~\ref{sec:beam},
  \item storage of the simulated events in a file which has the same format
  as the raw data,
  \item reconstruction of the simulated events with the same reconstruction
  chain as used for the real data,
  \item matching of the reconstructed to the simulated tracks based on
  the cluster positions.
\end{enumerate}

It should be underlined that only inelastic p+p interactions in the hydrogen
in the target cell were simulated
and reconstructed. Thus the Monte Carlo based corrections (see
Sec.~\ref{sec:analysis}) can be applied only for
inelastic events. The contribution of elastic events is removed by the event
selection cuts
(see Sec.~\ref{sec:event-cuts}), whereas the contribution of off-target
interactions is subtracted based on
the data (see Sec.~\ref{sec:off-target}).

  \section{Analysis procedure}\label{sec:analysis}

The analysis procedures consisted of the following steps:
\begin{enumerate}[(i)]
  \item applying event and track selection criteria,
  \item evaluation of the moments of distributions of quantities needed to
  calculate fluctuations
        (Eqs.~\ref{eq:omega},\ref{eq:delta},\ref{eq:sigma},\ref{eq:phi}),
  \item evaluation of corrections to the moments based on experimental data
  and simulations,
  \item calculation of the corrected fluctuations.
\end{enumerate}

Corrections for the following biases were evaluated and applied:
\begin{enumerate}[(i)]
  \item contribution of off-target interactions,
  \item losses of inelastic p+p interactions due to the trigger and the
  event and track selection criteria,
  \item contribution of particles other than primary charged hadrons,
  \item losses of primary charged hadrons due to the track selection
  criteria.
\end{enumerate}

The final results refer to charged hadrons produced in the
analysis acceptance in  inelastic
proton-proton interactions at 20, 31, 40, 80, and 158~\GeVc
beam momenta. Products of electromagnetic decays are included.
Products of weak decays and secondary interactions among the
tracks satisfying the selection criteria are corrected for.
The result is referred to as {\it accepted primary} hadrons.

The list of analyzed data sets together with statistics of
all recorded and selected events in target inserted and
target removed configurations is presented in Table~\ref{tbl:data-sets}.

\begin{table}[!htb]
  \caption {
    Data sets together with the statistics of events
    recorded and selected for the analysis
    in target inserted and
    target removed configurations.
  }
  \begin{center}
    \begin{tabular}{c|c|c|c c|c c}
      \noalign{\hrule height 0.03cm}\hline
      $p_{beam}$ & $\sqrt{s_{NN}}$ & \multirow{2}{*}{$y^{CM}_{beam}$} &
      \multicolumn{2}{|c|}{target inserted} & \multicolumn{2}{|c}{target
      removed} \cr
      [\GeVc] & [\GeV] & & all & selected & all & selected \cr \hline
      20 & 6.27 & 1.90 & 1\,324\,k & 255\,k & 122\,k & ~8\,k \cr
      31 & 7.62 & 2.10 & 3\,140\,k & 1058\,k & 332\,k & 35\,k \cr
      40 & 8.73 & 2.23 & 5\,226\,k & 2008\,k & 528\,k & 88\,k \cr
      80 & 12.32 & 2.57 & 4\,444\,k & 1791\,k & 458\,k & 88\,k \cr
      158 & 17.27 & 2.91 & 3\,538\,k & 1819\,k & 426\,k & 74\,k \cr
      \hline\noalign{\hrule height 0.03cm}
    \end{tabular}
  \end{center}
  \label{tbl:data-sets}
\end{table}

\subsection{Event selection criteria}\label{sec:event-cuts}

The following event selection criteria were applied to the events recorded
with the interaction trigger (Table~\ref{tbl:cuts}):
\begin{enumerate}[(i)]
  \item no off-time beam particle was detected within $\pm 1.5\ \mu$s around
  the trigger particle,
  \item the beam particle trajectory was measured in BPD-3 and at least one of BPD-1
  or BPD-2 detectors,
  \item there was at least one track reconstructed in the TPCs and fitted to
  the interaction vertex,
  \item events with a single, well measured positively charged track with
  absolute momentum close to the beam momentum ($p > p_{beam}$~-~1~\GeVc) were rejected.
  \item the vertex $z$ position (fitted using the beam and TPC tracks)
  was not farther away than 50~cm from the center of the LHT,
\end{enumerate}

The off-line (listed above) and on-line (the interaction trigger condition,
see Sec.~\ref{sec:beam}) event cuts
select well measured (cuts (i), (ii)) inelastic p+p interactions.
The background due to  elastic
interactions is removed (cuts (iii) and (iv)) and
the contribution of off-target interactions
is reduced (cut (v)) and was later subtracted
using data recorded in target
removed configuration.
The losses of inelastic interactions due to the event selection procedure
were corrected using a simulation (see below).

\subsection{Track selection criteria}\label{sec:track-cuts}

In order to select well-measured tracks of primary charged hadrons as well
as to reduce the
contamination of tracks from secondary interactions, weak decays and off-time
interactions the
following track selection criteria were applied (Table~\ref{tbl:cuts}):
\begin{enumerate}[(i)]
  \item the track momentum fit at the interaction vertex should have
  converged,
  \item the total number of reconstructed points on the track should be
  greater than~30,
  \item the sum of the number of reconstructed points in \mbox{VTPC-1} and
        \mbox{VTPC-2} should be greater than~15 or the number of
        reconstructed points in the GAP-TPC should be greater than~5,
  \item the distance between the track extrapolated to the interaction plane
        and the interaction point (impact parameter) should be smaller
        than 4~cm in the horizontal (bending) plane and 2~cm in the
        vertical (drift) plane,
  \item the track should be measured in a high ($\geq90\%$) TPC acceptance and
        tracking efficiency region (see Sec.~\ref{sec:acceptance}),
  \item tracks with energy loss and total momentum values characteristic for
        electrons were rejected.
  \item the transverse momentum was required to be less than 1.5~\GeVc.
\end{enumerate}

\begin{table}[!htb]
  \caption{
    Summary of event and track selection criteria used in the analysis.
  }
  \begin{tabular}{l|c|c|c}
    \cline{2-4}
    & standard cuts & loose cuts & tight cuts  \\ \hline
    T2 trigger & \multicolumn{3}{c}{applied}  \\ \cline{2-4}
    BPD & \multicolumn{3}{c}{applied} \\ \cline{2-4}
    off-time & $< \pm 1.5~\mu s$  & no cut & $< \pm 5~\mu s$ \\ \cline{2-4}
    fitted vertex z position & $\pm 50~cm$ & no cut & $\pm 10~cm$ \\ \cline{2-4}
    not elastic scatter & \multicolumn{3}{c}{applied} \\ \hline
    \hline
    total points & $\geq 30$ & no cut & $\geq 30$ \\ \cline{2-4}
    VTPC (GTPC) points & $\geq 15 (5)$ & $> 10(5)$ & $\geq 30(6)$ \\ \cline{2-4}
    $|b_x|$ & $\leq 4~cm$ & no cut & $\leq 2~cm$ \\ \cline{2-4}
    $|b_y|$ & $\leq 2~cm$ & no cut & $\leq 1~cm$ \\ \cline{2-4}
    $p_{T}$ & \multicolumn{3}{c}{$\le 1.5~\GeVc$} \\ \cline{2-4}
    $e^{\pm}$ & \multicolumn{3}{c}{applied} \\
    \hline\noalign{\hrule height 0.03cm}
  \end{tabular}
  \label{tbl:cuts}
\end{table}

\subsection{Determination of the analysis kinematical acceptance}\label{sec:acceptance}

\begin{figure*}[!htb]
  \centering
  \includegraphics[width=0.4\textwidth]{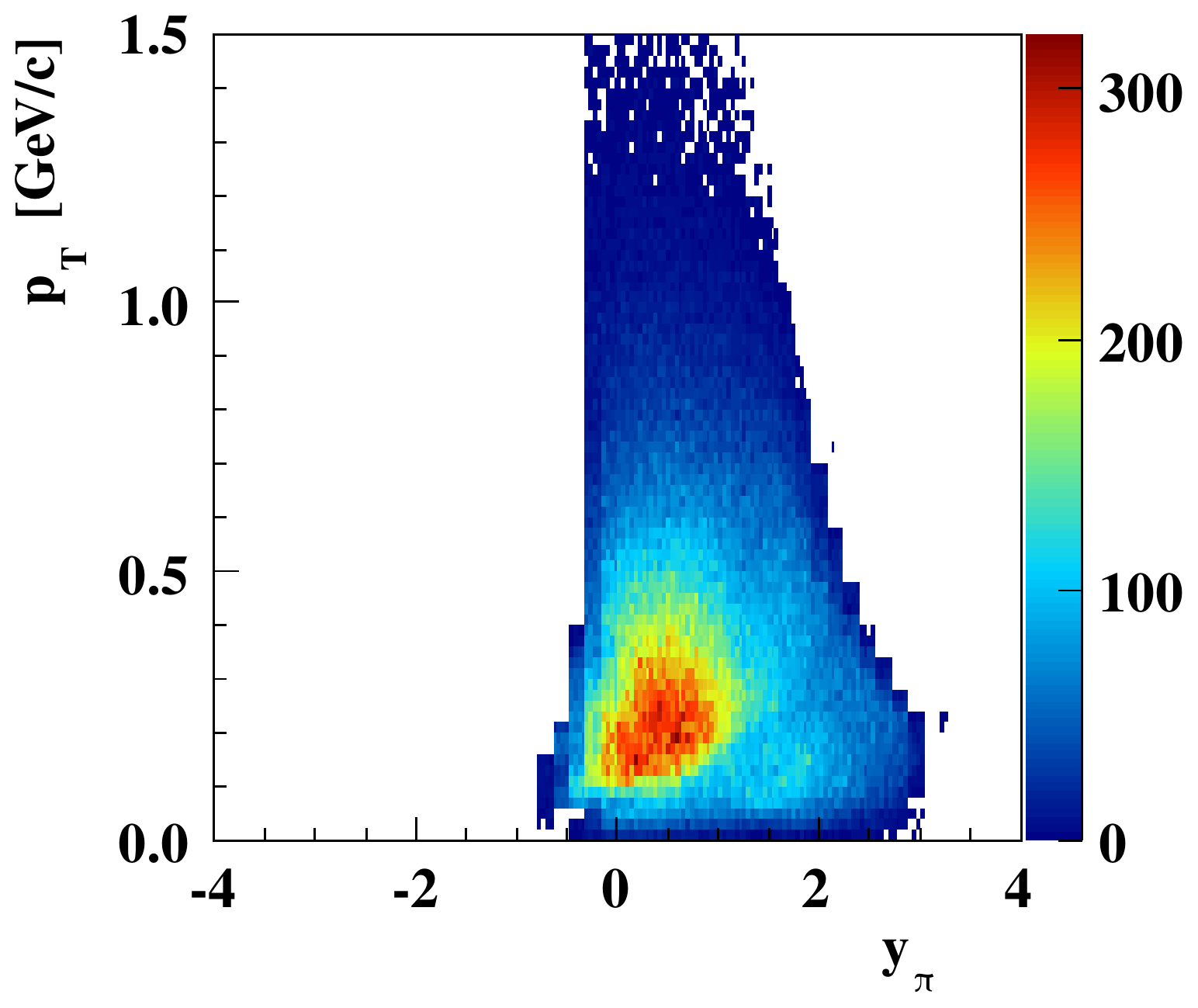} \hspace{1cm}
  \includegraphics[width=0.4\textwidth]{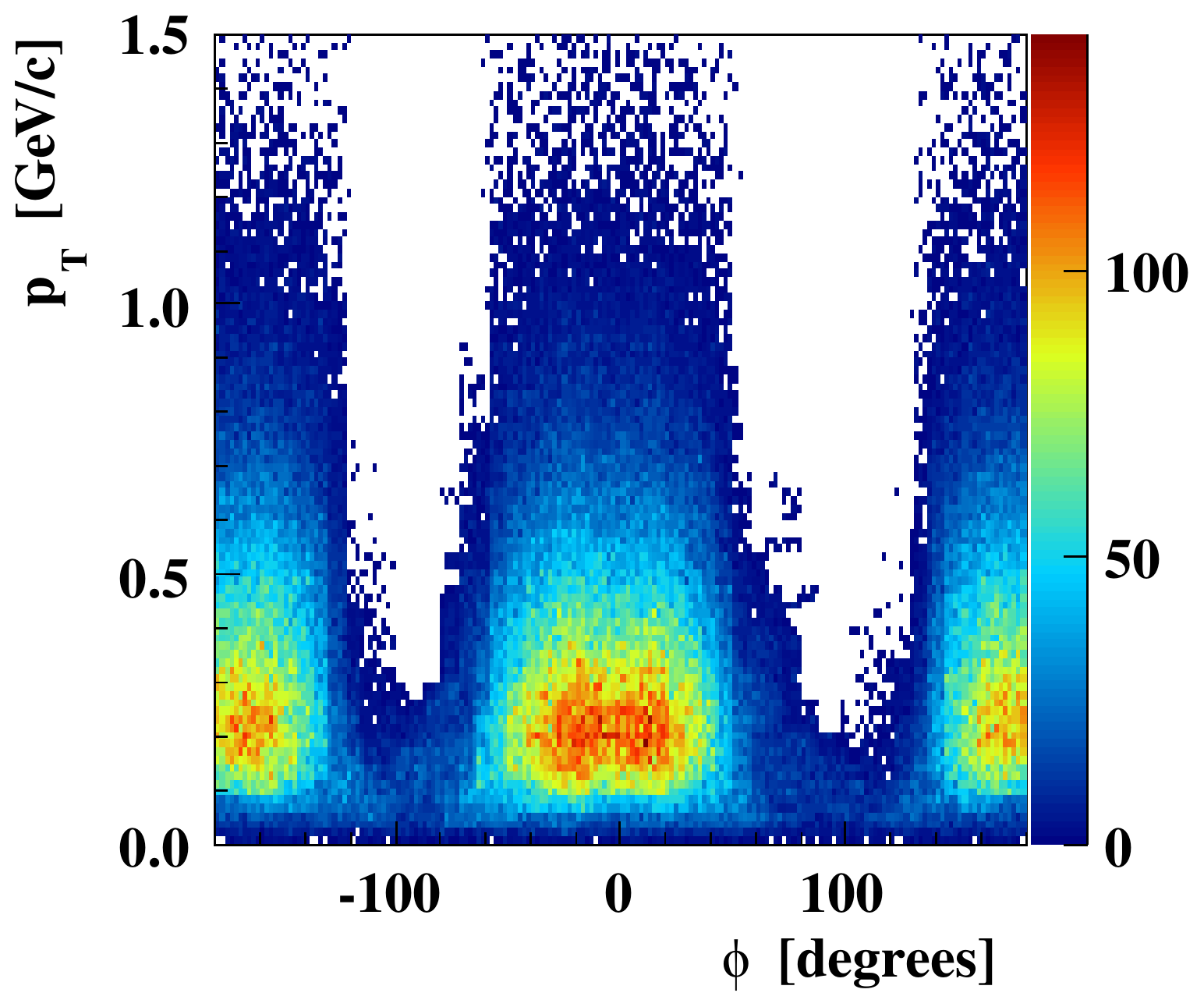} \\
  \includegraphics[width=0.4\textwidth]{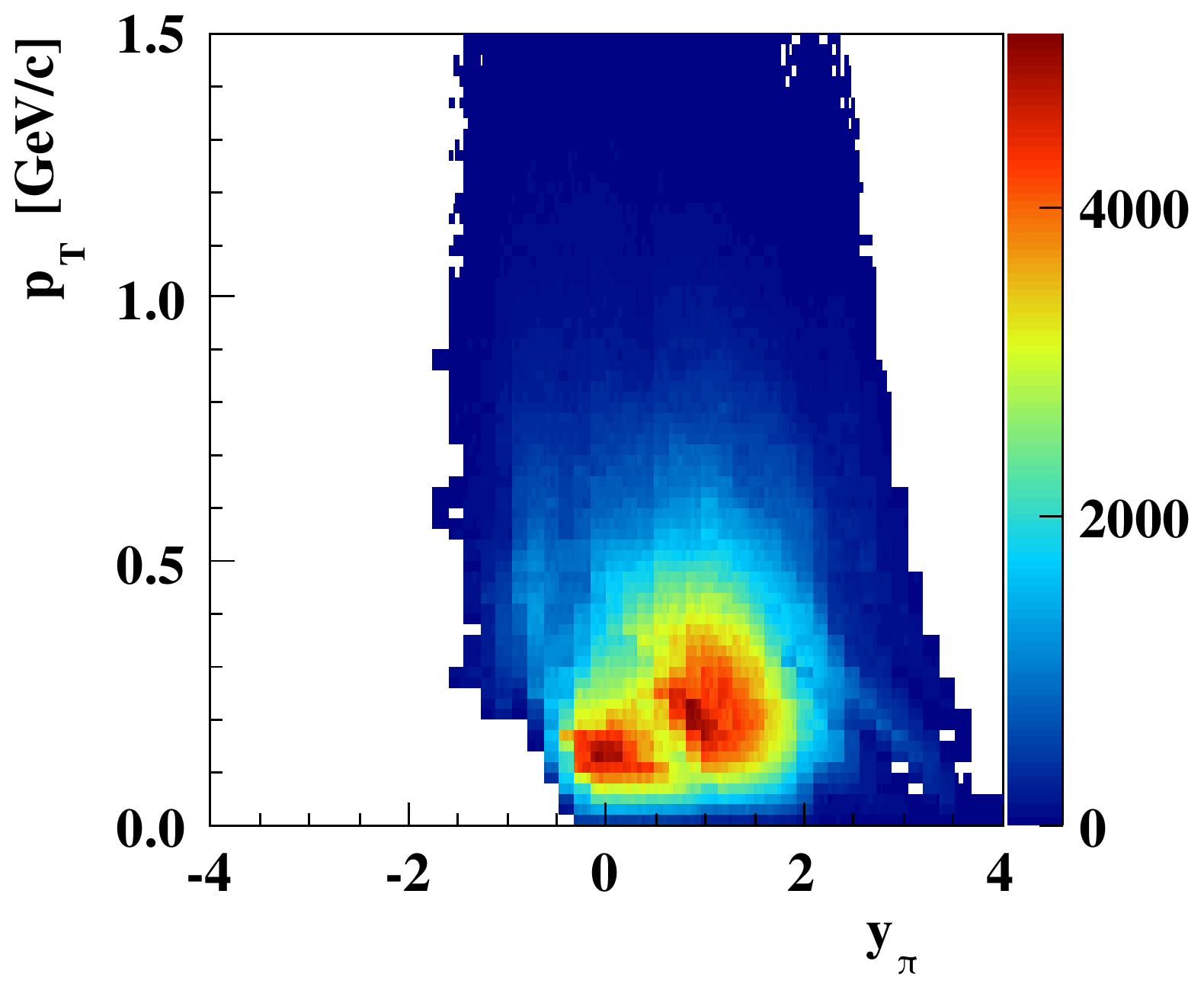} \hspace{1cm}
  \includegraphics[width=0.4\textwidth]{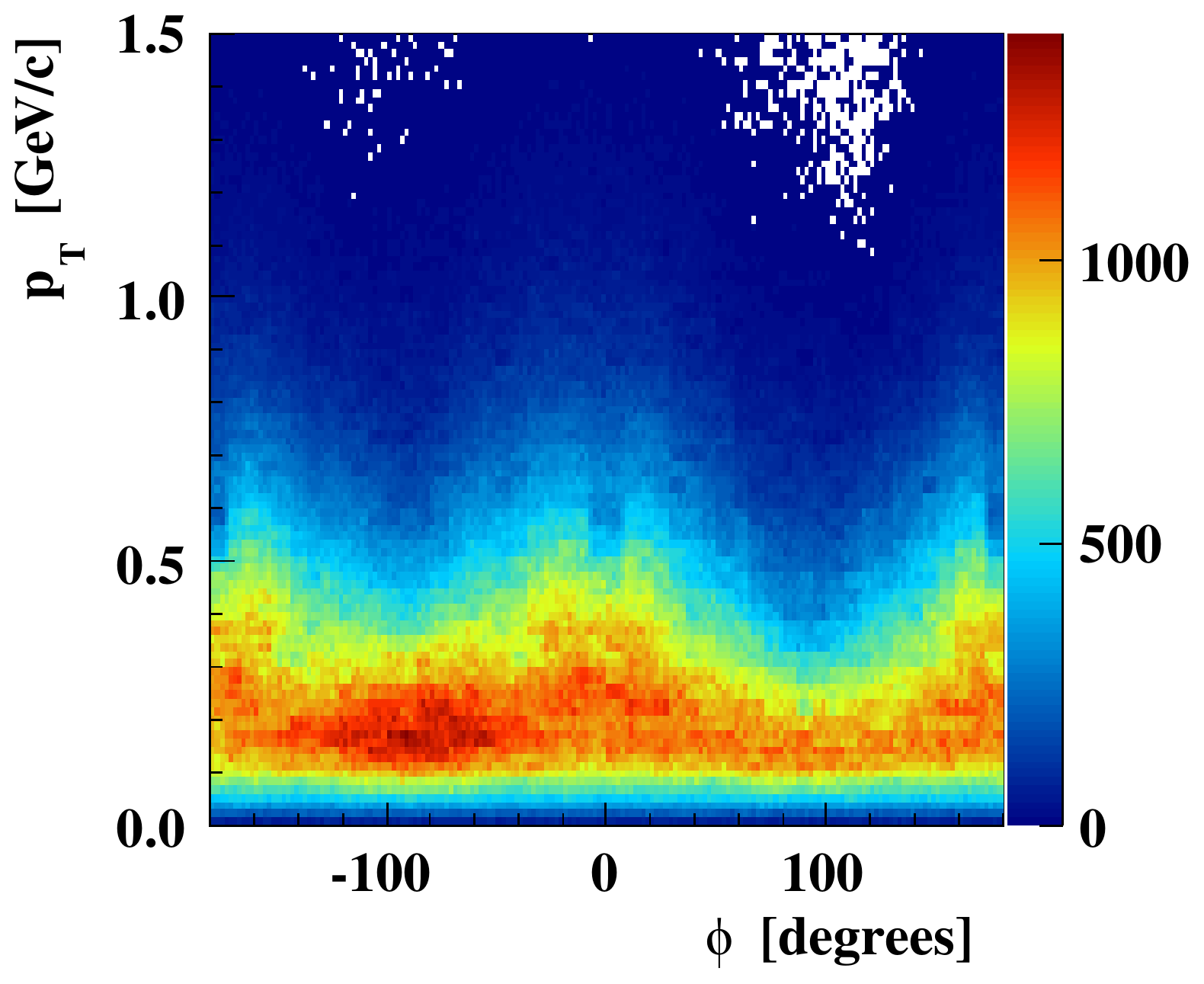}
  \caption[]{
    Population of all charged hadrons in the analysis acceptance
    used in this paper
    to study transverse momentum and multiplicity  fluctuations: the two top
    plots refer to 20~\GeVc and the two bottom plots to 158~\GeVc p+p interactions.
  }
  \label{fig:acc-ptflu}
\end{figure*}

The detection and reconstruction inefficiencies were corrected using the
simulation.
However, in order to limit the impact of possible inaccuracies of this
simulation, only regions were accepted where the reconstruction efficiency
(defined as the ratio of the number of reconstructed and matched
Monte Carlo tracks passing the track selection criteria to the
number of generated tracks) is greater than 90\%.
These regions were identified using a separate, statistically independent
simulation in three-dimensional bins of rapidity, azimuthal angle and
transverse momentum. The result is stored in the form of three-dimensional
tables~Ref.~\cite{edms-tables} where 0 signal bins excluded from
the acceptance and 1 those that are included.
The population of charged particles within this acceptance is
shown in Fig.~\ref{fig:acc-ptflu} for 20~\GeVc and 158~\GeVc p+p interactions.

\subsection{Data-based correction for off-target interactions}\label{sec:off-target}

\begin{figure}[!htb]
  \centering
  \includegraphics[width=0.8\columnwidth]{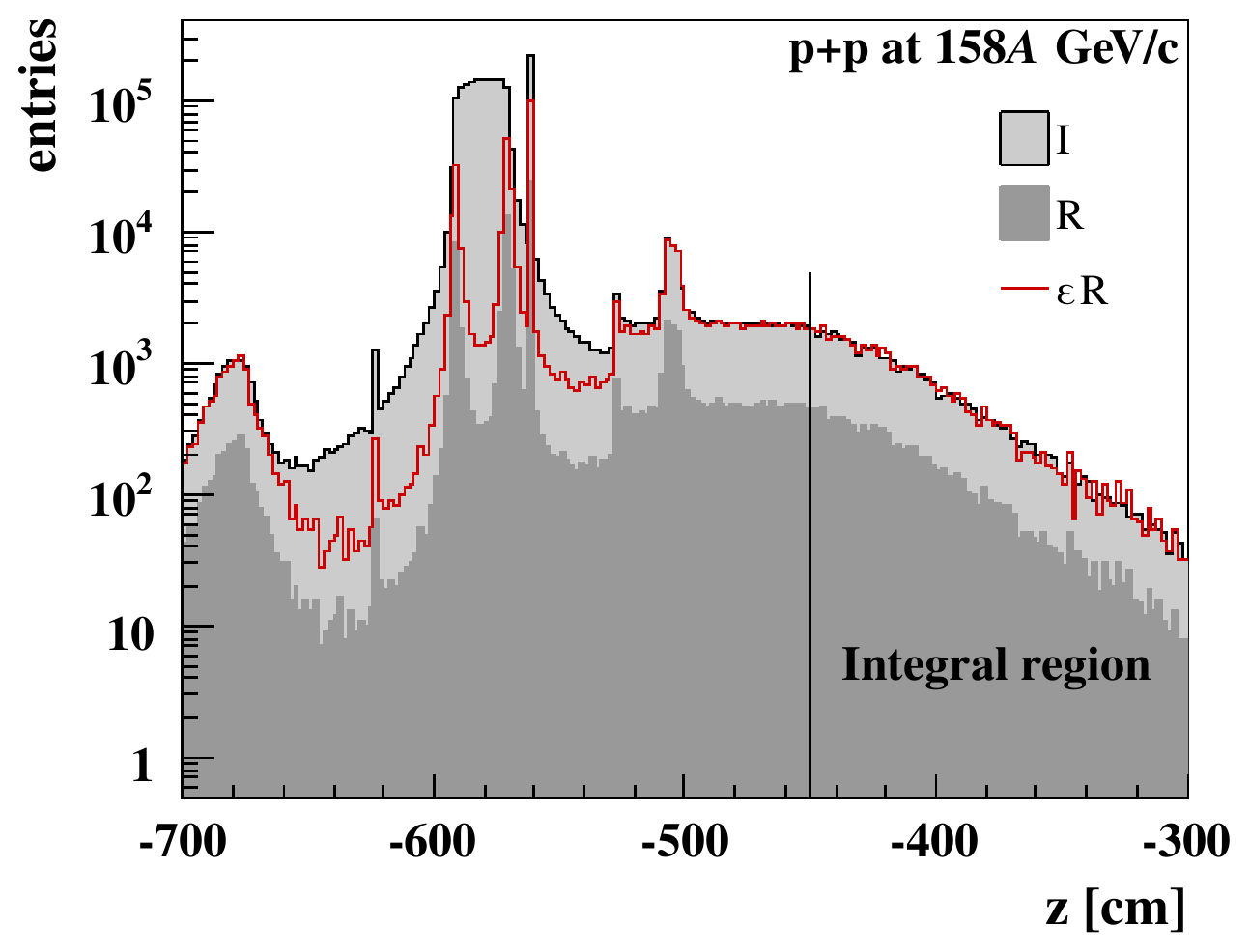}
  \caption{
    Distributions of the $z$ coordinate of the reconstructed interaction
    vertex for events recorded with the target inserted ($I$) and
    removed ($R$).
    The target removed distribution was normalized to the target inserted
    one in the region $z > -450$~cm.
  }
  \label{fig:corr-empty}
\end{figure}

The event quantities used to derive final fluctuation
measures were calculated
for events recorded in the LH filled (\textit{target inserted, I})
and removed (\textit{target removed, R}) configurations.
The latter data set represents
interactions with material downstream and upstream of the liquid
hydrogen (off-target interactions).
Then, in the absence of other corrections,
the corrected mean value of
the distribution of any quantity (denoted as $X$) was calculated as:
\begin{equation}\label{eq:off-target}
  \langle X \rangle = \frac{1}{N_{ev}^{I} - \epsilon \cdot N_{ev}^{R}} \left
  ( \sum\limits_{i=1}^{N_{ev}^{I}} X_{i}^{I} -
  \epsilon \cdot \sum\limits_{j=1}^{N_{ev}^{R}} X_{j}^{R} \right )~,
\end{equation}
where $N_{ev}$ denotes the number of events and $\epsilon$ is a normalization factor.
The value of $\epsilon$ was derived based on the distribution of the fitted $z$ coordinate
of the interaction vertex. All vertices far away from the target originate
from interactions
with the beam line and detector materials. Neglecting the beam attenuation
in the target one gets:
\begin{equation}\label{eq:epsilon}
  \epsilon =
  \frac{N_{ev}^{I}}{N_{ev}^{R}}\bigg|_{z>-450\mathrm{\ cm}}.
\end{equation}
Examples of distributions of the $z$ coordinate of the reconstructed
interaction vertex for events recorded
with the liquid hydrogen inserted and removed are shown in
Fig.~\ref{fig:corr-empty}.

\begin{figure*}[!htb]
  \centering
  \includegraphics[width=0.9\textwidth]{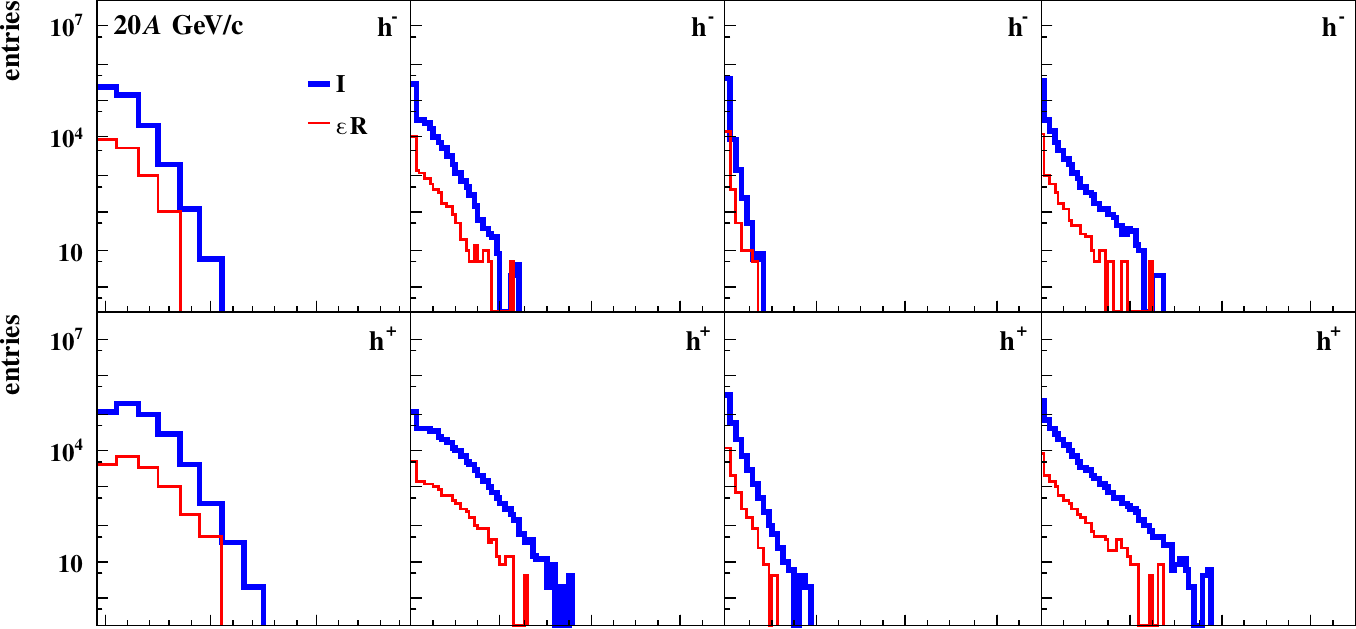} \\[0.25cm]
  \includegraphics[width=0.9\textwidth]{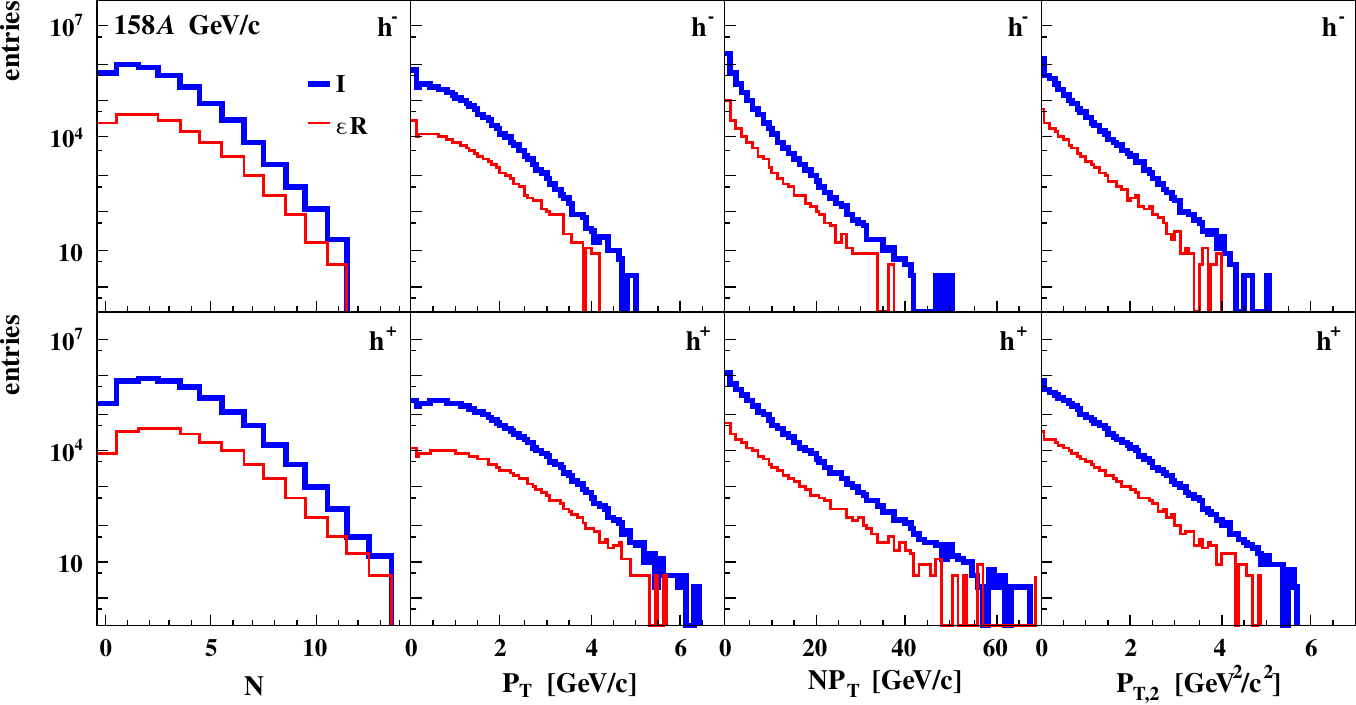}
  \caption{
    Examples of uncorrected distributions of
    event quantities for p+p interactions at 20 and 158~\GeVc beam
    for data recorded with the liquid hydrogen inserted (I) and
    removed (R).
    The spectra with the removed liquid hydrogen are multiplied
    by the $\epsilon$ factor (see Eq.~\ref{eq:epsilon}).
  }
  \label{fig:event-quantities}
\end{figure*}

\subsection{Simulation-based correction for other biases }

The correction for losses due to event and track selections,
reconstruction inefficiency and the interaction trigger, as well as for background of
non-primary charged hadrons was calculated using
the \EposLong~\cite{Werner:2008zza} event generator.
The simulated data were reconstructed with the standard \NASixtyOne procedure.
Tables of correction factors  were
calculated as the ratio of generated to reconstructed tracks.
The reconstructed tracks were required to pass
the event and track selection criteria.
The corrections were obtained in bins of $N$, $P_{T}$ and $P_{T,2} = \sum\limits_{i=1}^{N} p^{2}_{T_{i}}$ for
positively, negatively and all charged hadrons, separately.
The event quantity $P_{T,2}$ is needed to calculate $\omega[p_T]$ using only event quantities.
$\omega[p_T]$ appears in the  normalization factors $C_{\Delta}$ and $C_{\Sigma}$.
Thus there are three three-dimensional tables of correction factors.
Then for a given charge selection an event $i$ with
$N$, $P_{T}$ and $P_{T,2}$
is weighted with the correction factor $c_{i}$ from the table of
corrections for this charge selection and from the bin which corresponds
to $N$, $P_{T}$ and $P_{T,2}$.
Thus, in the absence of off-target interactions, the corrected
mean value of a quantity $X$ is:
\begin{equation}\label{eq:corr-mc}
  \langle X \rangle = \frac{1}{ M_{ev} } \left (
  \sum\limits_{i=1}^{ N_{ev} }
  c_{i}~X_{i}
  \right )~,
\end{equation}
where $M_{ev} = \sum\limits_{i=1}^{ N_{ev} } c_{i} $~.

\subsection{The final correction}\label{sec:final-corr}

The final results were obtained by combining the data-based
correction for off-target interactions with the Monte Carlo
based correction for other biases.  It was calculated as:
\begin{equation}\label{eq:corr-final}
  \langle X \rangle = \frac{1}{ M_{ev}^I - \epsilon~M_{ev}^R }
  \left (
  \sum\limits_{i=1}^{ N_{ev}^I } c_{i}~X_{i}^I -
  \epsilon \cdot \sum\limits_{j=1}^{ N_{ev}^R }
  c_{j}~X_{j}^R
  \right )~.
\end{equation}

\begin{figure*}[!htb]
  \centering
  \includegraphics[width=0.9\textwidth]{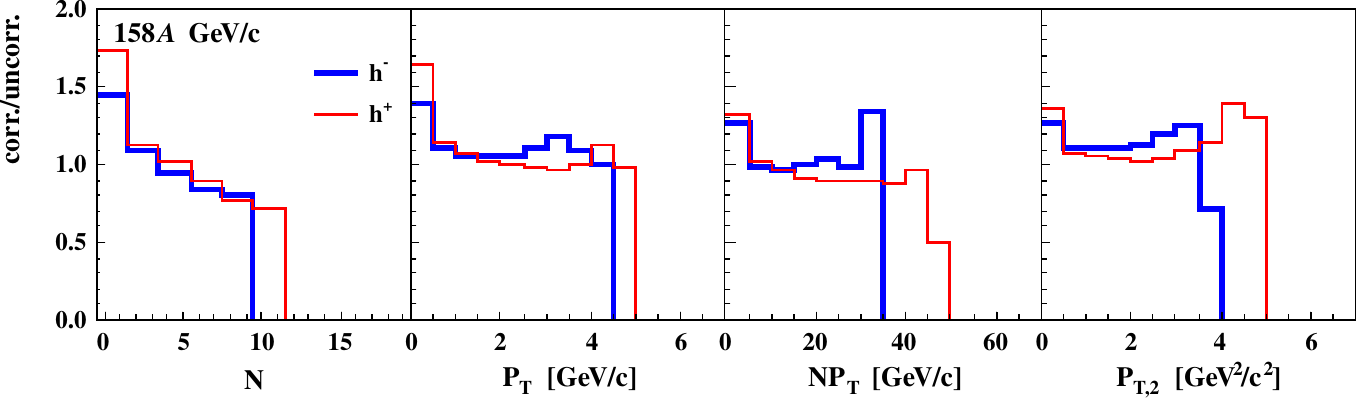}
  \caption{
    (Color online)
    Examples of ratios of corrected to uncorrected distributions of
    event quantities for  p+p interactions at 158~\GeVc.
  }
  \label{fig:corr-ratio}
\end{figure*}

In order to illustrate
the correction procedure and its impact on the results
selected  distributions of the relevant event quantities,
$N$, $P_T$, $NP_T$ and $P_{T,2}$,
and results on
$\Delta[P_{T},N]$, $\Sigma[P_{T},N]$, $\Phi_{p_{T}}$ and $\omega[N]$
obtained at the subsequent stages of the procedure are presented
and discussed.

Figure~\ref{fig:event-quantities}
shows uncorrected distributions of the event quantities
for data recorded with proton beams at 20~\GeVc and 158~\GeVc
with the liquid hydrogen inserted and removed.
The spectra with the removed liquid hydrogen are multiplied
by the $\epsilon$ factor defined in Eq.~\ref{eq:epsilon}.
The distributions with the LH inserted and removed have similar shape.
The normalized spectra for the LH removed are about 10 times
lower than the ones for the LH inserted.
Thus the correction for the off-target interactions is expected
to be small (see below).

Figure~\ref{fig:corr-ratio} presents
the ratio of fully corrected (see Eq.~\ref{eq:corr-final})
to uncorrected distributions of the
event quantities for  p+p interactions at 158~\GeVc
for positively and negatively charged particles, separately.
The ratio varies significantly from about 0.5 to about 1.7.

\begin{figure*}[!htb]
  \centering
  \includegraphics[width=0.9\textwidth]{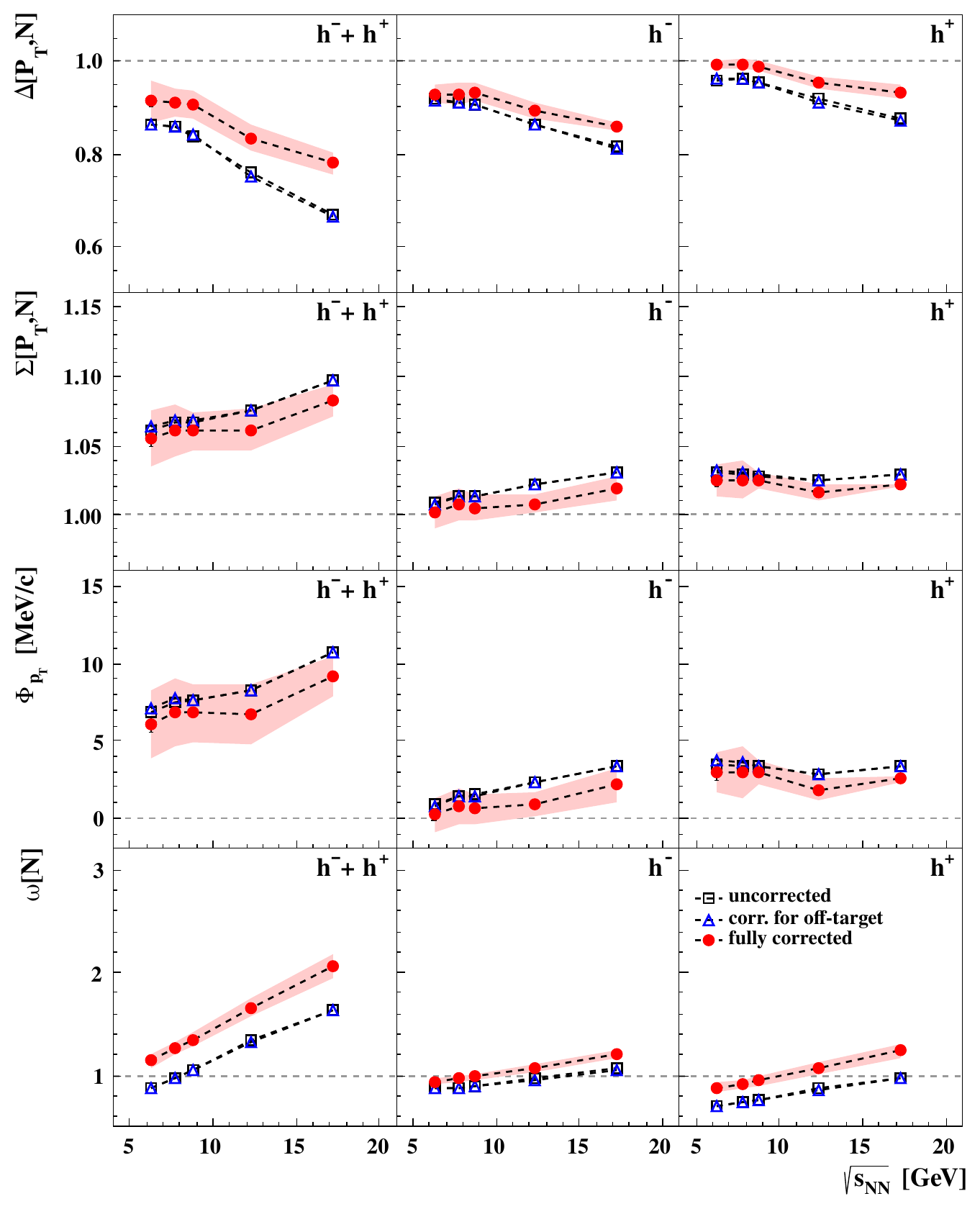}
  \caption{
    (Color online)
    Results on $\Delta[P_{T},N]$, $\Sigma[P_{T},N]$, $\Phi_{p_{T}}$ and $\omega[N]$
    as a function of collision energy before
    corrections (open squares), corrected for contributions of off-target interactions (open circles)
    and additionally corrected for all other experimental biases, see Sec.~\ref{sec:final-corr}
    (filled circles). Statistical uncertainties (mostly invisible) are shown by vertical bars,
    systematic uncertainties by shaded bands.
  }
  \label{fig:corr-effect}
\end{figure*}

\begin{figure*}[!htb]
  \centering
  \includegraphics[width=0.4\textwidth]{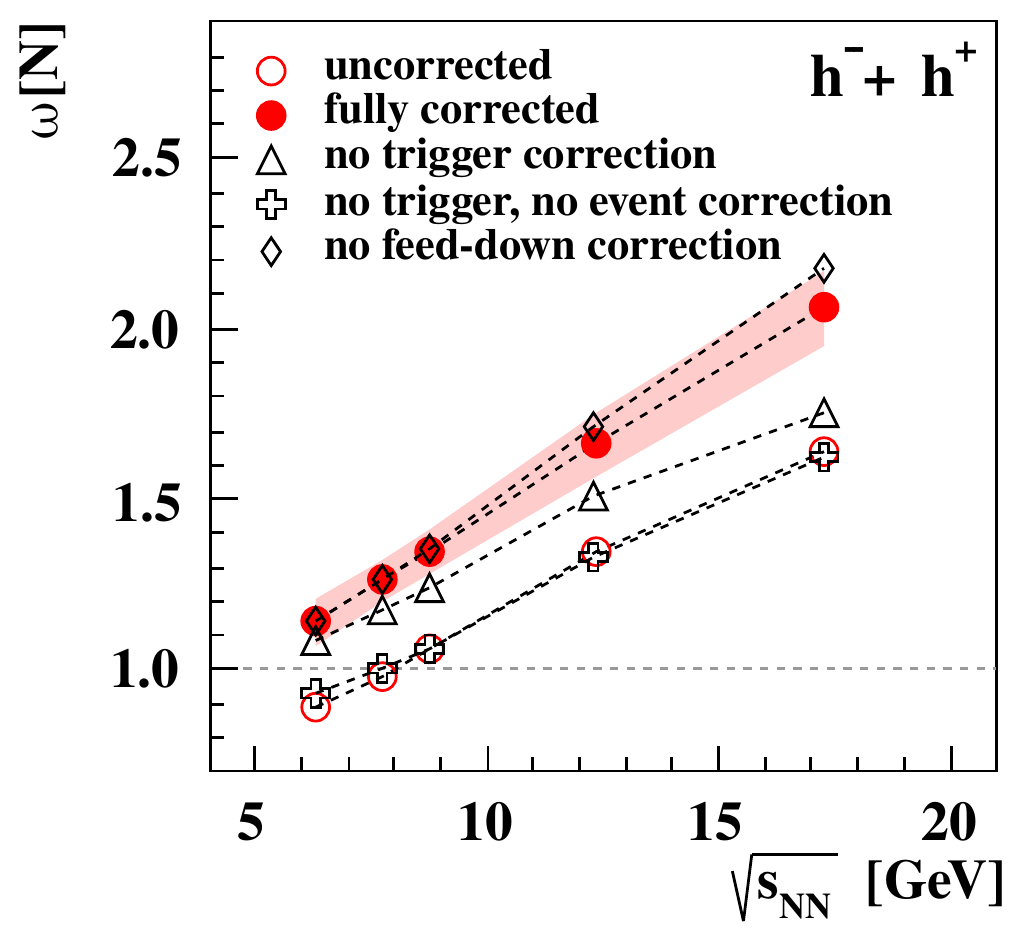}\hspace{1cm}
  \includegraphics[width=0.4\textwidth]{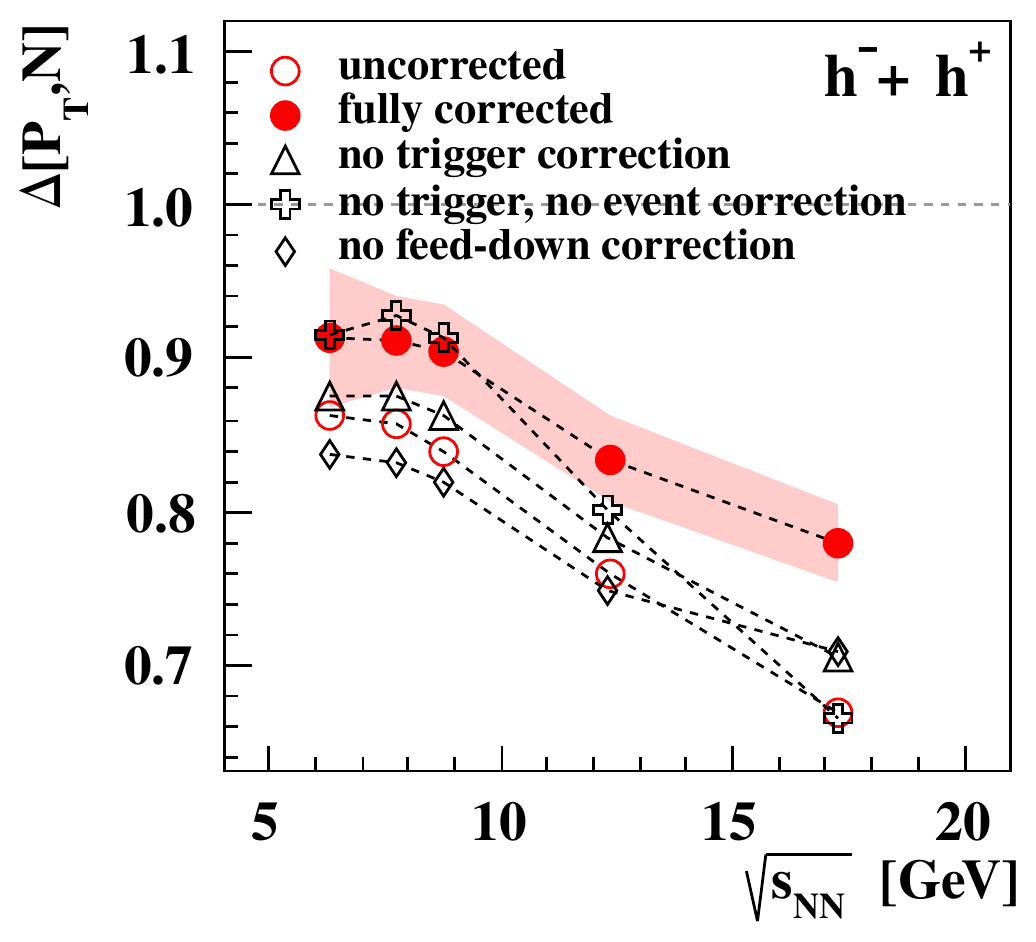}
  \caption{
    (Color online)
    Results on $\omega[N]$ and $\Delta[P_{T},N]$ as a function of collision energy
    for all charged hadrons
    after all corrections (full circles), without the correction for
    the trigger bias (upward pointing triangles), without corrections for the
    trigger bias and the off-line event selection (crosses), without correction for
    feed-down (diamonds) and uncorrected (open circles).
    Statistical uncertainties (mostly invisible) are shown by vertical bars,
    systematic uncertainties by shaded bands.
  }
  \label{fig:event_cuts}
\end{figure*}

Results for uncorrected,
corrected only for the off-target interactions and
fully corrected data are shown in Fig.~\ref{fig:corr-effect}.
Statistical and systematic uncertainties (see below) of the fully corrected
points are also plotted for comparison.
The corrections for off-target interactions only weakly change the results.
The corrections for the remaining experimental biases have significant
impact in particular on results for $\omega[N]$ and $\Delta[P_T,N]$.
It is mostly due to the requirement of a well fitted interaction vertex
as well as corrections for the trigger bias and
the off-line selection of events.
This is illustrated in Fig.~\ref{fig:event_cuts} where
the collision energy dependence of $\omega[N]$ and $\Delta[P_T,N]$
for fully corrected data, uncorrected for the trigger bias,
uncorrected for the trigger bias and for the off-line event selection
as well as fully uncorrected data are presented.
In addition, the results with all corrections but the correction for
the contribution of non-primary tracks (feed-down) are shown.
The corrections

\subsection{Statistical uncertainties}

The statistical uncertainties of $\Phi_{p_{T}}$, $\Delta[P_{T},N]$ and
$\Sigma[P_{T},N]$ were estimated as follows.
The whole sample of events was divided into 30 independent sub-samples. The
values of $\Phi_{p_{T}}$,
$\Delta[P_{T},N]$, and $\Sigma[P_{T},N]$ were evaluated for each sub-sample
separately (following all the
procedures described above, including target removed and Monte Carlo corrections)
and the dispersions
($D_{\Phi}$, $D_{\Delta}$, and $D_{\Sigma}$) of the results were then
calculated. The statistical uncertainty
of $\Phi_{p_{T}}$ ($\Delta[P_{T},N]$ or $\Sigma[P_{T},N]$) is taken to be
equal to $D_{\Phi}/\sqrt{30}$
($D_{\Delta}/\sqrt{30}$ or $D_{\Sigma}/\sqrt{30}$).

For each beam momentum, 5 million events were generated and reconstructed,
several times more than the recorded experimental data. Therefore
statistical uncertainties arising from the event statistics of the
simulation were neglected.

\subsection{Systematic uncertainties}

Systematic uncertainties were estimated by changing:
\begin{enumerate}[(i)]
  \item event and track selection criteria: {\it tight} and {\it loose} cuts
        (see Table~\ref{tbl:cuts}) and,
  \item model chosen for the simulation: \EposLong~\cite{Werner:2008zza} and
        \VenusLong~\cite{Werner:1993uh,Werner:1990aa}.
\end{enumerate}
To estimate the contribution of each source, the complete analysis was repeated
under these different conditions.

The uncertainties of corrections for the  event selection and feed-down are expected to be mostly due to uncertainties in modelling of p+p interactions,
whereas the uncertainties of the remaining corrections (e.g. for the track reconstruction
inefficiency) are expected to be mostly due to imperfectness of the detector simulation.
Total systematic uncertainties were calculated by adding in quadrature uncertainties calculated in (i) and (ii).
  \section{Results}\label{sec:results}

The results shown in this section refer to primary {\it accepted} hadrons produced in
all inelastic p+p interactions. The accepted hadrons are hadrons produced
within the kinematical acceptance selected for the analysis~\cite{edms-tables},
see also Sec.~\ref{sec:analysis}.
The results  are corrected for event and track losses due to
detector inefficiencies, selection criteria and the interaction trigger, as well as
contamination of  tracks from weak decays and secondary interactions
and leptons from primary interactions. The correction procedure
is described in detail in Sec.~\ref{sec:analysis}.
Results are {\it not} corrected for the kinematic acceptance. This
acceptance should be taken into
account when the data are compared with models.
Table~\ref{tbl:multiplicities} shows mean multiplicities of negatively and positively charged hadrons within the kinematical acceptance selected for the analysis in
this paper.

\begin{table*}[!htb]
  \centering
  \small
  \caption{
   Mean multiplicities of negatively and positively charged hadrons produced
   in inelastic p+p interactions at 20, 31, 40, 80 and 158~\GeVc
   in the kinematical acceptance used in this paper and in the NA49 publications.
   For comparison also mean multiplicity of $\pi^-$ mesons in full phase space
   as obtained by \NASixtyOne~\cite{Abgrall:2013pp_pim} is shown in the second
   column.
  }
  \label{tbl:multiplicities}
  \begin{tabular}{c | c | c c | c c | c c | c c }
    \noalign{\hrule height 0.03cm}\hline
      \multicolumn{2}{c}{} &
      \multicolumn{2}{|c|}{\NASixtyOne} &
      \multicolumn{2}{|c|}{NA49-$N$ \cite{Anticic:2008aa}} &
      \multicolumn{2}{|c|}{NA49-$M$ \cite{Alt:2007jq}} &
      \multicolumn{2}{|c}{NA49-$B$ \cite{Alt:2007jq}} \cr
    \cline{3-10}
      \multicolumn{2}{c}{} &
      \multicolumn{2}{|p{2.5cm}|}{\scriptsize kinematical acceptance used in the analysis (see Sec.~\ref{sec:acceptance}) } &
      \multicolumn{2}{|p{2.5cm}|}{\scriptsize narrow $\phi$ acc. common for all energies; \newline $1.1 < y_{\pi} < 2.6$} &
      \multicolumn{2}{|p{2.5cm}|}{\scriptsize no VTPC-1-only tracks; \newline $1.1 < y_{\pi} < y_{beam}$} &
      \multicolumn{2}{|p{2.5cm}}{\scriptsize no VTPC-1-only tracks; \newline $0 < y_{\pi} < y_{beam}$} \cr
    \hline
    $p_\mathrm{beam}$[\GeVc] &
    $\langle \pi^{-}\rangle$ \cite{Abgrall:2013pp_pim}&
    $\langle h^{-}\rangle$ & $\langle h^{+}\rangle$ &
    $\langle h^{-}\rangle$ & $\langle h^{+}\rangle$ &
    $\langle h^{-}\rangle$ & $\langle h^{+}\rangle$ &
    $\langle h^{-}\rangle$ & $\langle h^{+}\rangle$ \cr
    \hline
    20  & 1.01 & 0.34 & 0.91 & 0.04 & 0.18 & 0.10 & 0.32 & 0.18 & 0.50 \cr
    31  & 1.31 & 0.51 & 1.14 & 0.06 & 0.20 & 0.16 & 0.41 & 0.29 & 0.66 \cr
    40  & 1.48 & 0.64 & 1.30 & 0.07 & 0.21 & 0.21 & 0.48 & 0.38 & 0.76 \cr
    80  & 1.94 & 1.04 & 1.78 & 0.10 & 0.23 & 0.41 & 0.74 & 0.66 & 1.11 \cr
    158 & 2.44 & 1.49 & 2.26 & 0.15 & 0.25 & 0.68 & 1.09 & 1.05 & 1.56 \cr
    \hline\noalign{\hrule height 0.03cm}
  \end{tabular}
\end{table*}

\begin{figure*}[!htb]
  \centering
  \includegraphics[width=.9\textwidth]{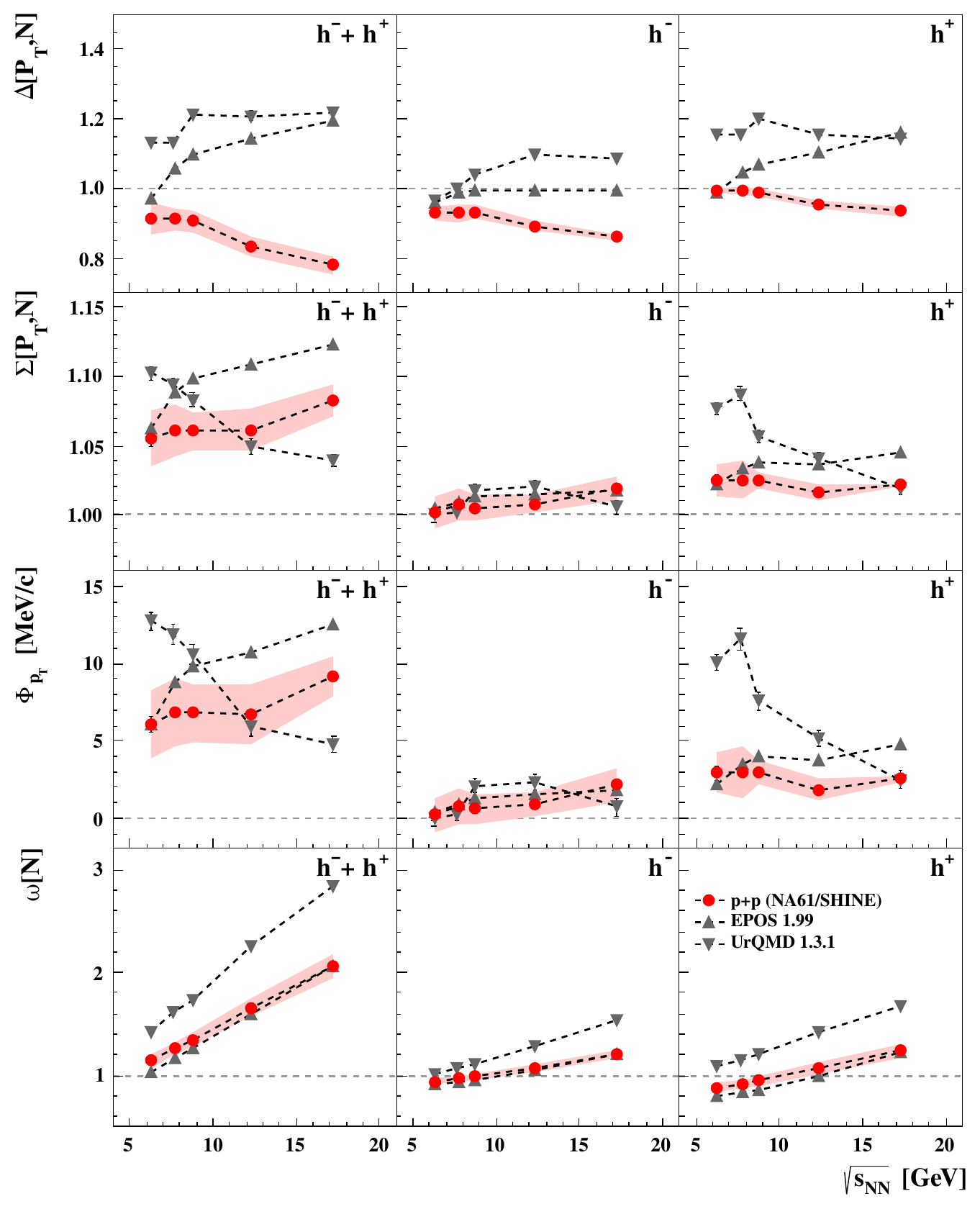}
  \caption{
    (Color online)
    Collision energy dependence of $\Delta[P_{T},N]$, $\Sigma[P_{T},N]$, $\Phi_{p_{T}}$
    and $\omega[N]$ for inelastic p+p interactions.
    The NA61/SHINE measurements (filled circles) are compared with predictions of the
    \EposLong~(upward pointing triangles) and
    UrQMD (downward pointing triangles) models (both in the NA61/SHINE acceptance).
    Statistical uncertainties (mostly invisible) are shown by vertical bars,
    systematic uncertainties by shaded bands.
  }
  \label{fig:results}
\end{figure*}

Figure~\ref{fig:results} shows the results on
$\Delta[P_{T},N]$, $\Sigma[P_{T},N]$, $\Phi_{p_T}$ and $\omega[N]$
calculated separately for
all charged, negatively charged, and positively charged hadrons produced
in inelastic p+p interactions at  20--158~\GeVc beam momentum.

First, one observes that the results, in general,
significantly deviate from the reference values
$\Delta[P_{T},N] = \Sigma[P_{T},N] = 1$ and $\Phi_{p_T} =0$~\MeVc
which are expected in case of independent particle production.

Second,
the results for the three charge selections show differences.
The effect of long-range correlations , e.g., caused by conservation laws,
decreases with decreasing fraction of accepted particles.
In particular, many charged hadrons come from decays of resonances into
two or more hadrons, e.g., $\rho \rightarrow \pi^+ + \pi^-$.
Positively charged hadrons are mostly  $\pi^{+}$ mesons and protons.
Less of them come from resonance decays  into two or more positively charged
hadrons. The majority of negatively charged hadrons are negatively charged pions
and only a small fraction comes from resonance decays  into two or more
negatively charged hadrons.
Thus the correlations due to resonance decay kinematics decrease
from all charged hadrons to positively and negatively ones.
Other sources of correlations like quantum statistics and Coulomb interactions
are also expected to impact fluctuations differently for different charge selections.

Third, the collision energy dependence of $\Delta[P_{T},N]$
and  $\Sigma[P_{T},N]$ are opposite.
With increasing collision energy
$\Delta[P_{T},N]$ decreases, whereas $\Sigma[P_{T},N]$ increases.
The largest deviations from unity for both $\Delta[P_{T},N]$ and
$\Sigma[P_{T},N]$ are observed for all charged hadrons at 158~\GeVc.
Note, that at this energy the fraction of accepted particles is the largest.
  \section{Comparison with models and central Pb+Pb collisions}
\label{sec:models}

Figure~\ref{fig:results} shows
a comparison of the measured fluctuations
with predictions of two commonly used
models:~\EposLong~\cite{Werner:2008zza}  and
UrQMD~\cite{Bleicher:1999xi}.
The predictions were calculated for the \NASixtyOne acceptance~\cite{edms-tables}.

The \EposLong predictions agree quantitatively with results on $\omega[N]$,
They exhibit the same trend with increasing collision energy as the experimental results on
$\Sigma[P_{T},N]$ and $\Phi_{p_T}$, but there is qualitative disagreement with
results on $\Delta[P_{T},N]$.
Note that the \EposLong model agrees reasonably well with single
particle spectra of identified hadrons produced in the same
inelastic p+p interactions~\cite{Pulawski:2015tka}.

Essentially all UrQMD predictions disagree with the data.
In many cases even qualitative disagreement is observed.
Note, that UrQMD also fails to describe single particle spectra
in p+p reactions~\cite{Pulawski:2015tka}.
These disagreements are probably due to problems in modelling of
hadron production via resonance decays~\cite{Vovchenko:2014vda}.

\NASixtyOne records data at beam momenta per nucleon identical
to or close to those at which NA49~\cite{Afanasev:1999iu}
performed measurements of central Pb+Pb collisions.
This allows for a direct study of the system size
dependence of various hadron production properties in the SPS
energy range.

The \NASixtyOne results presented in this paper are obtained
in a phase space acceptance~\cite{edms-tables} which is larger
than the acceptances used by NA49 to obtain the corresponding results
for central Pb+Pb collisions~\cite{Alt:2006jr,Alt:2007jq,Anticic:2003fd,Anticic:2008aa}.
Thus, in order to compare the \NASixtyOne measurements
with the NA49 data, the more restrictive NA49 cuts were applied to
the \NASixtyOne data.

The narrowest acceptance (referred as to the NA49-$N$ acceptance ) was
used in the NA49 study of collision energy dependence of
(transverse momentum)-multiplicity fluctuations~\cite{Anticic:2008aa}.
The NA49-$N$ acceptance is limited to the rapidity range
$1.1 < y_{\pi} < 2.6$, where $y_{\pi}$ is the rapidity calculated in the cms
assuming the pion mass, and selects particles in a common narrow azimuthal angle wedge
at all beam momenta.

\begin{figure*}
  \centering
  \includegraphics[width=0.9\textwidth]{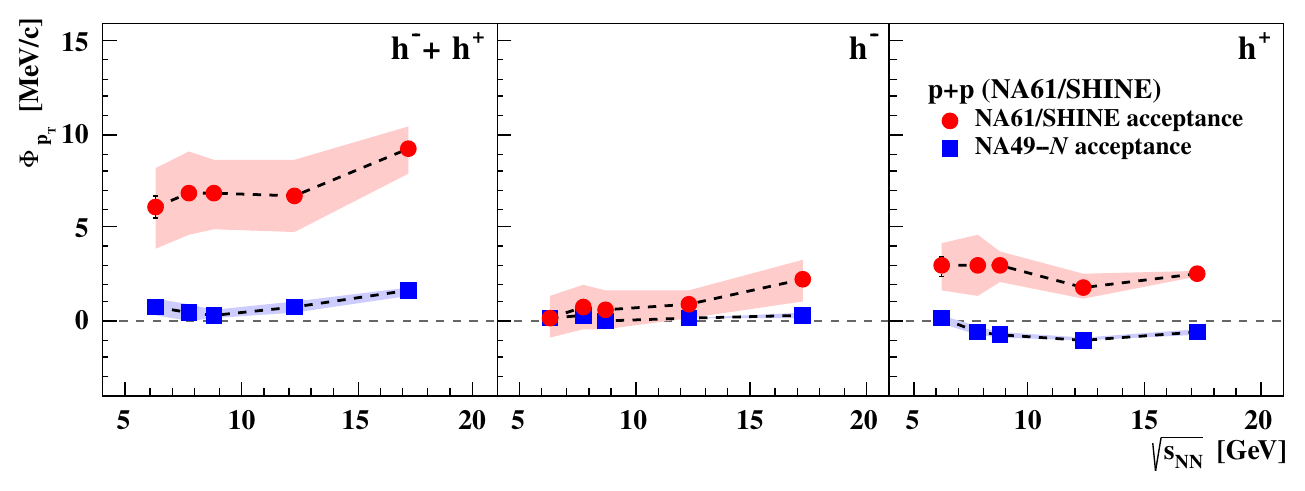}
  \caption[]{
    (Color online) \NASixtyOne results on $\Phi_{p_T}$ vs
    collision energy for inelastic p+p interactions within the full
    \NASixtyOne acceptance (see Fig.~\ref{fig:results}) and within
    the NA49-$N$ acceptance (see Ref.~\cite{Anticic:2008aa}).
    Statistical errors (mostly invisible) are shown by vertical bars,
    systematic uncertainties by shaded bands.
  }
  \label{fig:Phipt_na49na61}
\end{figure*}

Figure~\ref{fig:Phipt_na49na61} shows the \NASixtyOne results on  $\Phi_{p_T}$ in
inelastic p+p interactions within the full \NASixtyOne acceptance and within the
NA49-$N$ acceptance. As expected, the restriction of the acceptance strongly
reduces the values of the fluctuation measure.

\begin{figure*}
  \centering
  \includegraphics[width=0.9\textwidth]{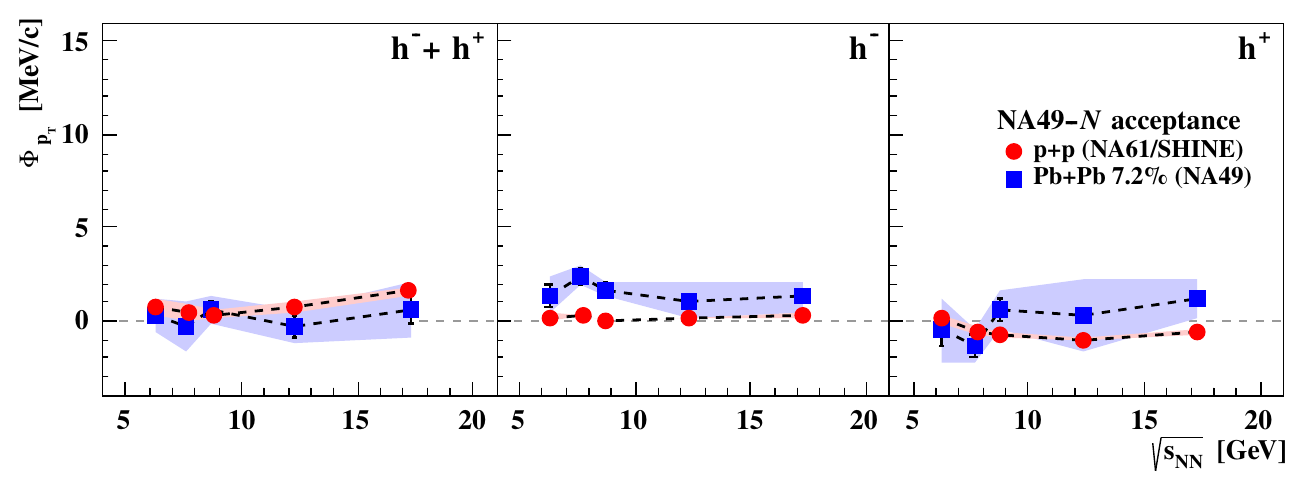}
  \caption[]{
    (Color online) $\Phi_{p_T}$ versus collision energy for inelastic p+p (\NASixtyOne)
    interactions and the 7.2\% most central Pb+Pb
    (NA49~\cite{Anticic:2008aa}) collisions in the NA49-$N$ acceptance.
    Statistical errors are shown by vertical bars, systematic uncertainties by shaded bands.
  }
  \label{fig:Phipt_energy}
\end{figure*}

Figure~\ref{fig:Phipt_energy} presents a comparison of $\Phi_{p_T}$ for
inelastic p+p (\NASixtyOne) interactions and the 7.2\% most central
Pb+Pb (NA49) collisions within the NA49-$N$ acceptance.
No significant difference is observed between the results for the two
reactions. Moreover, neither shows a structure in the collision
energy dependence which could be attributed to the onset of deconfinement
or the critical point.

\begin{figure*}
  \centering
  \includegraphics[width=0.9\textwidth]{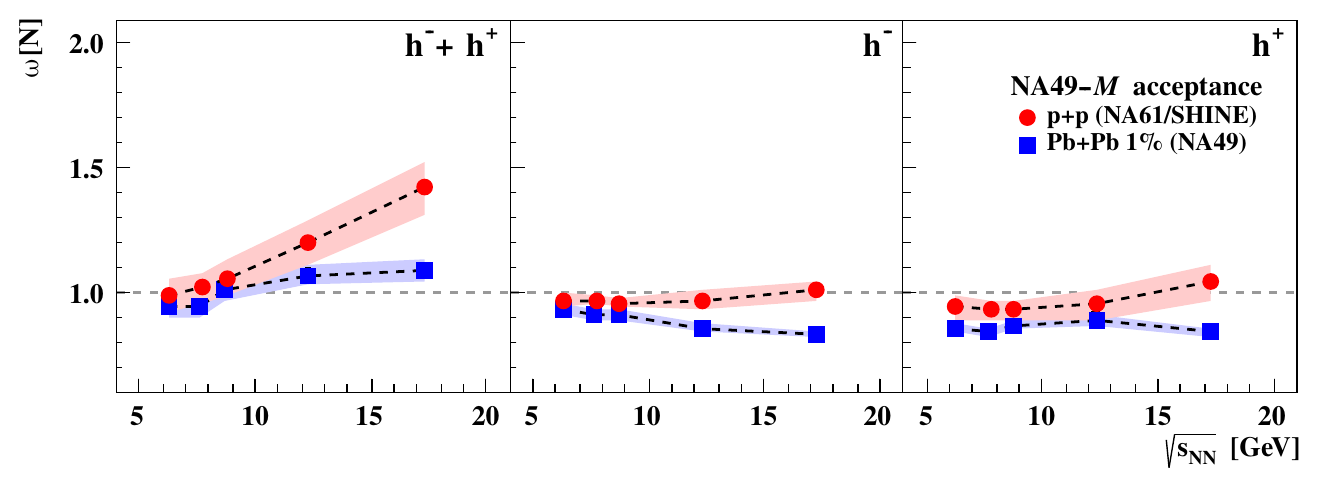}
  \includegraphics[width=0.9\textwidth]{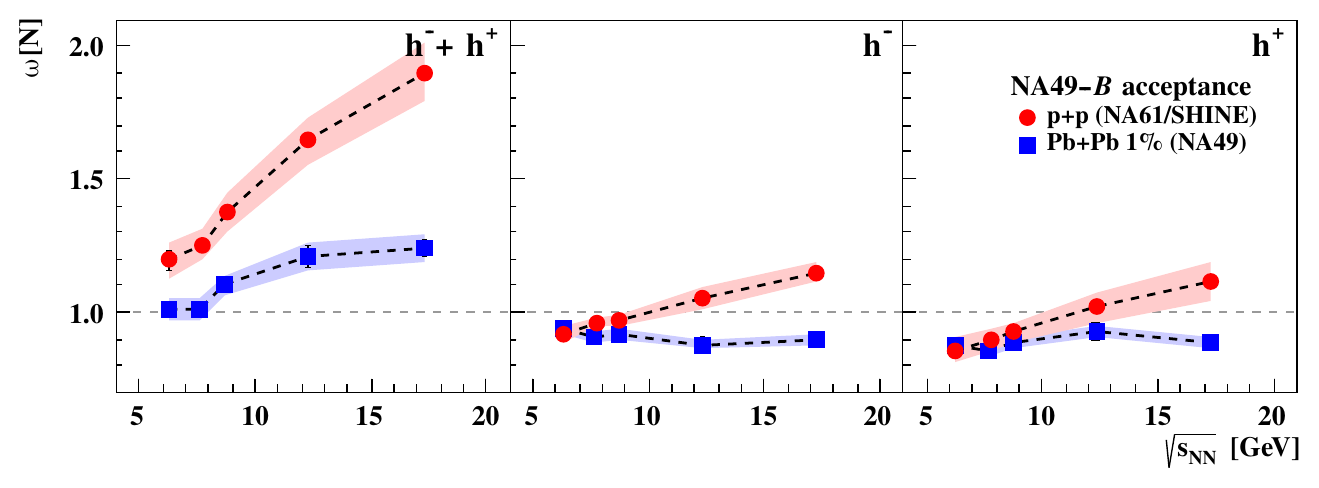}
  \caption[]{
    (Color online)
    Collision energy dependence of the scaled variance of the multiplicity distribution
    for inelastic p+p (\NASixtyOne)
    interactions and the 1\% most central Pb+Pb (NA49~\cite{Alt:2007jq}) collisions in
    the NA49-$M$ ($top$) and NA49-$B$ ($bottom$) acceptances (see text for details).
    Statistical errors (mostly invisible) are shown by vertical bars,
    systematic uncertainties by shaded bands.
  }
  \label{fig:omega}
\end{figure*}

Figure~\ref{fig:omega} shows the collision energy dependence of
the scaled variance of the multiplicity distributions for
inelastic p+p (\NASixtyOne) interactions and the 1\% most central
Pb+Pb (NA49) collisions within the NA49-$M$~(top row) and NA49-$B$~(bottom row)
acceptances~\cite{Alt:2007jq}.
The NA49 medium (NA49-$M$) acceptance includes all particles well measured
by the NA49 detector within the rapidity range
$1.1 < y_{\pi} < y_{beam}$
and the NA49 broad acceptance (NA49-$B$) extends the range
to $0 < y_{\pi} < y_{beam}$.
Significant differences are observed between the results for p+p and Pb+Pb
reactions, see below for a discussion.
However, neither shows a structure in the collision
energy dependence which could be attributed to the onset of deconfinement
or the critical point.

The scaled variance is significantly larger for inelastic p+p interactions
at 158~\GeVc than for the 1\% most central Pb+Pb collisions at 158\AGeVc.
In the following this difference will be discussed within
the Wounded Nucleon Model (WNM)~\cite{Bialas:1976ed} and
the Statistical Model (SM)~\cite{Fermi:1950jd} of particle production.
These models are selected  because they are simple and play a special
role in analyzing the physics of heavy ion collisions.
The discussion will be focused on the results for negatively charged
hadrons as they are weakly influenced by decays of
resonances~\cite{Begun:2006uu}.

\begin{figure*}
  \centering
  \includegraphics[width=0.4\textwidth]{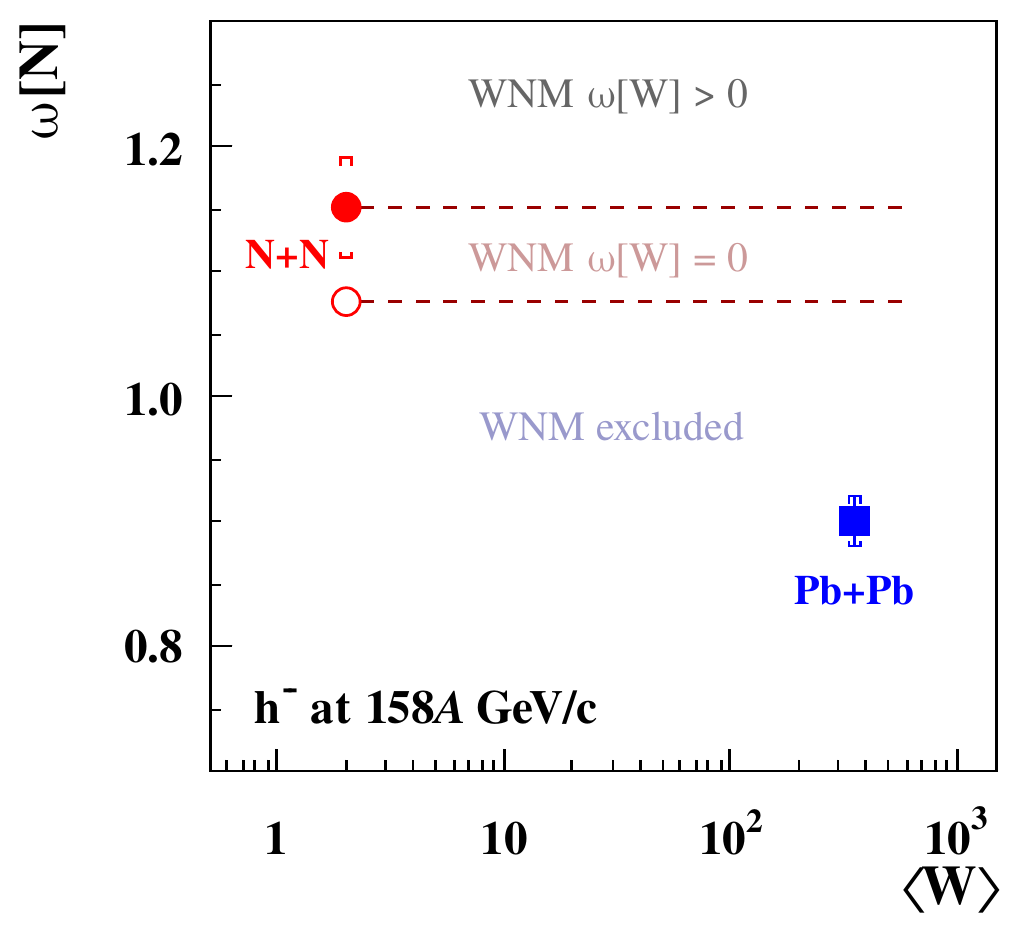}\hspace{1cm}
  \includegraphics[width=0.4\textwidth]{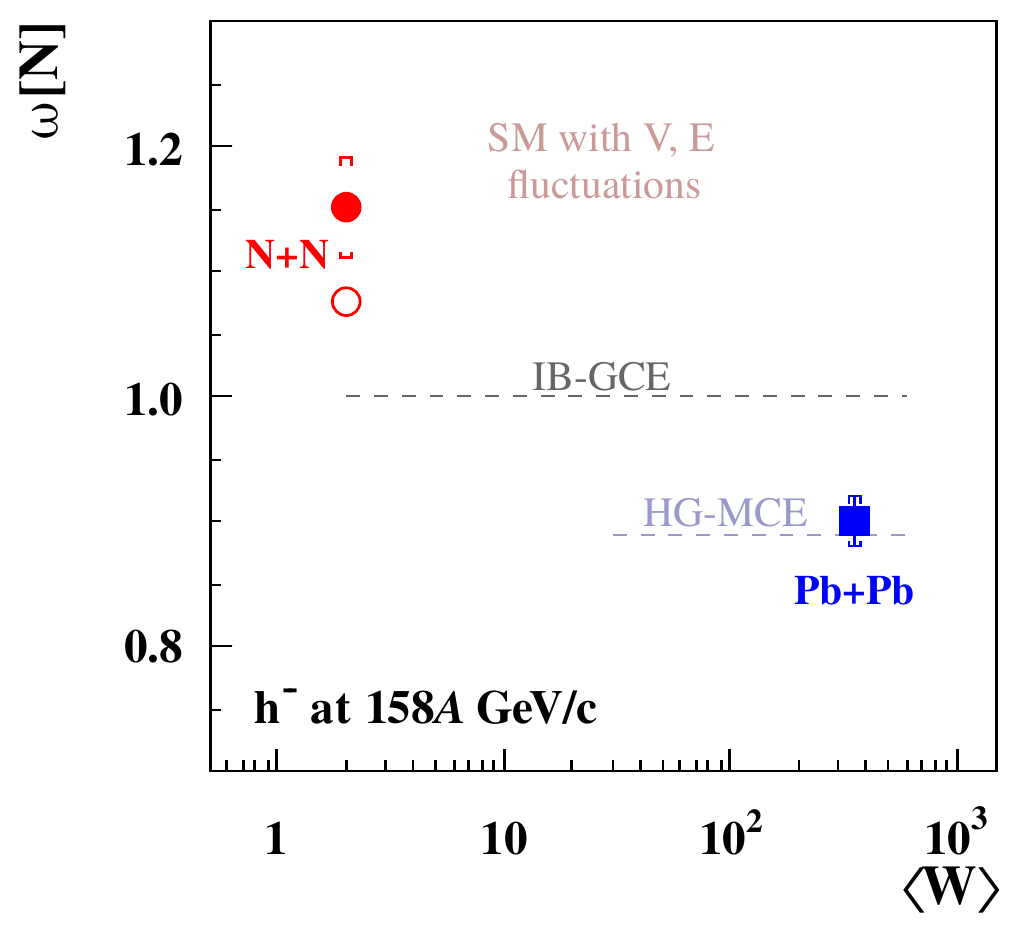}
  \caption[]{
    (Color online)
    Scaled variance of the multiplicity distribution of negatively charged
    hadrons as a function of the mean number of wounded nucleons.
    Results for inelastic p+p (\NASixtyOne) interactions (filled circles)
    and the 1\% most central Pb+Pb (NA49~\cite{Alt:2007jq}) collisions (squares) in
    the NA49-$B$  acceptances are shown together with the predictions of the
    Wounded Nucleon Model ({\it left}) and the statistical model ({\it right}) (see text for explanations).
  }
  \label{fig:omega_wnm}
\end{figure*}

The Wounded Nucleon Model~\cite{Bialas:1976ed} assumes that particle production in nucleon-nucleon
and nucleus-nucleus collisions is an incoherent superposition
of particle production from wounded nucleons (nucleons
which interacted inelastically and whose number is calculated using
the Glauber approach).
Properties of wounded nucleons are assumed to be independent of the size of the colliding
nuclei, e.g. they are the same in p+p and Pb+Pb collisions at the same
collision energy per nucleon.
The scaled variance of the multiplicity distribution of produced particles
calculated within the WNM reads~\cite{Gorenstein:2011vq}:
\begin{equation}
 \omega[N] = \omega^*[N] + \langle N \rangle / \langle W \rangle \cdot \omega[W]~,
\label{eq:wnm:varN}
\end{equation}
where $W$ is the number of wounded nucleons and
 $\omega^*[N]$ denotes the scaled variance calculated for any fixed value
of $W$. The number of wounded nucleons (protons) in p+p interactions is
fixed to two and the measured scaled variance for p+p interactions can
be used as $\omega^*[N]$.  The second component of the sum Eq.~\ref{eq:wnm:varN}
vanishes or is positive.
Consequently the WNM predicts that the scaled variance in nucleus-nucleus
collisions has to be equal or larger than the one in inelastic
proton-proton interactions. The \NASixtyOne and NA49 results
presented in Fig.~\ref{fig:omega_wnm} clearly
contradict this prediction.

Note that at SPS energies  multiplicity
distributions of negatively charged hadrons in inelastic p+p, n+p and
n+n interactions are different~\cite{Golokhvastov:2001ei}.
The largest difference is observed between the distributions in
p+p and n+n interactions. Thus the prediction of the WNM for Pb+Pb
collisions which takes
into account the isospin effects estimated using the \Epos
model ranges between the results for p+p (the measured \NASixtyOne data) and
n+n interactions (estimated based on the \Epos predictions and the \NASixtyOne data).
This range is bounded in Fig.~\ref{fig:omega_wnm}~({\it left})
by the horizontal dashed lines.

The Statistical Model of multi-particle production was initiated
by Fermi in 1950~\cite{Fermi:1950jd}.
Its basic assumption states that all possible micro-states of
the macroscopic system created in a collision are equally probable.
For a non-interacting (ideal) gas of Boltzmann  particles
in the grand canonical ensemble (IB-GCE) the multiplicity distribution
is Poissonian ($\omega[N] = 1$) independent of the (fixed) system volume
(upper dashed line in Fig.~\ref{fig:omega_wnm}~({\it right})).
Resonance decays and Bose effects increase the scaled variance
from 1 to 1.1,
whereas the conservation laws reduce it if the system volume is sufficiently
large~\cite{Begun:2006uu}. The combined effect is demonstrated by the
lower dashed line in Fig.~\ref{fig:omega_wnm}~({\it right}).
In fact the NA49 result for the 1\% most central Pb+Pb collisions at 158\AGeVc
is well described by the hadron gas model in the micro--canonical
ensemble (HG-MCE)~\cite{Begun:2006uu}.
Within the statistical models
a scaled variance significantly larger than one as measured in inelastic p+p
interactions at high collision energies (top SPS and higher)
can be understood as a result of
volume and/or energy fluctuations~\cite{Begun:2008fm}.

Multiplicity and transverse momentum fluctuations quantified using
strongly intensive measures were studied in a number of
theoretical papers.
In particular, the influence of resonance decays~\cite{Mrowczynski:1999vi,Gorenstein:2013nea},
quantum statistics~\cite{Mrowczynski:1999vi,Gorenstein:2013iua}
and a dependence of the mean transverse momentum
on multiplicity~\cite{Mrowczynski:1999vi,Gorenstein:2013nea} was considered.
These studies are important for a qualitative understanding of
experimental data and predictions of complicated Monte Carlo models.
However, the obtained results cannot be directly compared to the measurements as
they did not include important effects. In particular, the limited
experimental acceptance is difficult to take into account in
simple models.
  \section{Summary}
\label{sec:summary}

This paper presents measurements of
multiplicity and transverse momentum fluctuations
of negatively, positively and all charged hadrons produced in
inelastic p+p interactions at 20, 31, 40, 80 and 158~\GeVc beam momentum.
Values for the scaled variance of multiplicity
distributions and three strongly intensive measures of multiplicity--transverse
momentum fluctuations $\Delta[P_{T},N]$, $\Sigma[P_{T},N]$ and $\Phi_{p_T}$
were obtained.
These results were calculated in the \NASixtyOne acceptance which has to be taken into
account in a quantitative comparison with models and other  results.
For the first time the results on fluctuations are fully
corrected for experimental biases, in particular, for the losses of inelastic events
due to the trigger and analysis event selections as well as for the contamination of
particles from weak decays and secondary interactions.

The measurements of multiplicity and transverse momentum fluctuations significantly deviate
from expectations for independent particle production
($\Delta[P_{T},N] = \Sigma[P_{T},N] = 1$, $\Phi_{p_T} =0$~\MeVc).
They also depend on the charges of the selected hadrons.
The scaled variances of the multiplicity distributions increase with increasing
collision energy and for all charged hadrons at 158~\GeVc reach the value of 2.

The string-resonance Monte Carlo models \Epos and \Urqmd do not describe
the data well. In several cases the collision energy dependence predicted by the models
even shows a trend qualitatively different from the measurements.

The scaled variance of multiplicity distributions is significantly higher in
inelastic p+p interactions than in the 1\% most central Pb+Pb collisions measured
by NA49 at the same energy per nucleon.
The largest difference is observed at 158\AGeVc.
This result is in qualitative disagreement with the predictions of the
Wounded Nucleon Model.
The low level of multiplicity fluctuations measured in central Pb+Pb collisions is
well reproduced by the statistical model in the micro-canonical ensemble
formulation. Within the statistical framework the enhanced multiplicity
fluctuations in inelastic p+p interactions can be interpreted as due to
event-by-event fluctuations of the fireball energy and/or volume.


  \begin{acknowledgements}
    We would like to thank the CERN PH, BE and EN Departments for the
strong support of NA61.

This work was supported by the Hungarian Scientific Research Fund
(grants OTKA 68506 and 71989), the J\'anos Bolyai Research Scholarship
of the Hungarian Academy of Sciences, the Polish Ministry of Science
and Higher Education (grants 667\slash N-CERN\slash2010\slash0,
NN\,202\,48\,4339 and NN\,202\,23\,1837), the Polish National Center
for Science (grants~2011\slash03\slash N\slash ST2\slash03691,
2013\slash11\slash N\slash ST2\slash03879,
2014\slash13\slash N\slash ST2\slash02565,
2014\slash14\slash E\slash ST2\slash00018
and
2015\slash18\slash M\slash ST2\slash00125),
the Foundation for Polish Science --- MPD program, co-financed by the
European Union within the European Regional Development Fund, the
Federal Agency of Education of the Ministry of Education and Science
of the Russian Federation (SPbSU research grant 11.38.242.2015), the
Russian Academy of Science and the Russian Foundation for Basic
Research (grants 08-02-00018, 09-02-00664 and 12-02-91503-CERN), the
Ministry of Education, Culture, Sports, Science and Tech\-no\-lo\-gy,
Japan, Grant-in-Aid for Sci\-en\-ti\-fic Research (grants 18071005,
19034011, 19740162, 20740160 and 20039012), the German Research
Foundation (grant GA\,1480/2-2), the EU-funded Marie Curie Outgoing
Fellowship, Grant PIOF-GA-2013-624803, the Bulgarian Nuclear
Regulatory Agency and the Joint Institute for Nuclear Research, Dubna
(bilateral contract No. 4418-1-15\slash 17), Ministry of Education and
Science of the Republic of Serbia (grant OI171002), Swiss
Nationalfonds Foundation (grant 200020\-117913/1), ETH Research
Grant TH-01\,07-3 and the U.S.\ Department of Energy.

  \end{acknowledgements}

  \bibliographystyle{includes/na61Utphys}
  \bibliography{includes/na61References}

\end{document}